\documentclass[twocolumn,aps,prl,amssymb,amsmath,superscriptaddress]{revtex4-2}

\usepackage[utf8]{inputenc}
\usepackage{amsmath}
\usepackage{bm}
\usepackage{color}
\usepackage{amssymb}
\usepackage{mathtools}
\usepackage{graphicx}
\usepackage[dvipsnames]{xcolor}
\usepackage{comment}
\usepackage{caption}
\DeclareCaptionLabelSeparator{vertbar}{ \pmb{$\vert$} }
\DeclareCaptionFormat{smallfont}
{\small #1#2#3}
\makeatletter
\renewcommand{\fnum@figure}{\textbf{Fig.~\thefigure}}

\usepackage{cmbright}

\usepackage{xr} 
\newcommand{\figref}[1]{Fig. \ref{#1}}

\newcommand{\Vbias}{{V_\text{bias}}}

\DeclarePairedDelimiterX{\braket}[2]{\langle}{\rangle}{#1 \delimsize \vert #2}

\begin{document}

\title{An electrical molecular motor driven by angular momentum transfer}
\author{Julian Skolaut}
\affiliation{Institute for Quantum Materials and Technology, Karlsruhe Institute of Technology, D-76021 Karlsruhe, Germany}
\author{Štěpán Marek}
\affiliation{Department of Condensed Matter Physics, Faculty of Mathematics and Physics, Charles University, Ke Karlovu 5, 121 16, Praha 2, Czech Republic}
\author{Nico Balzer}
\affiliation{Institute of Nanotechnology, Karlsruhe Institute of Technology (KIT), D-76021 Karlsruhe, Germany}
\author{María Camarasa-Gómez}
\affiliation{Departamento de Polímeros y Materiales Avanzados: Física, Química y Tecnología, Facultad de Química, UPV/EHU,
 20018 Donostia-San Sebastián, Spain}
\affiliation{Centro de Física de Materiales CFM/MPC (CSIC-UPV/EHU), 20018 Donostia-San Sebastián, Spain}
\author{Jan Wilhelm}
\affiliation{Institute  of  Theoretical  Physics and Regensburg Center for Ultrafast Nanoscopy (RUN),  University  of  Regensburg,   D-93053  Regensburg,  Germany}
\author{Jan Lukášek}
\affiliation{Institute of Nanotechnology, Karlsruhe Institute of Technology (KIT), D-76021 Karlsruhe, Germany}
\author{Michal Valášek}
\affiliation{Institute of Nanotechnology, Karlsruhe Institute of Technology (KIT), D-76021 Karlsruhe, Germany}
\author{Lukas Gerhard}
\affiliation{Institute for Quantum Materials and Technology, Karlsruhe Institute of Technology, D-76021 Karlsruhe, Germany}
\author{Ferdinand Evers}
\affiliation{Institute  of  Theoretical  Physics and Regensburg Center for Ultrafast Nanoscopy (RUN),  University  of  Regensburg,   D-93053  Regensburg,  Germany}
\author{Marcel Mayor}
\affiliation{Institute of Nanotechnology, Karlsruhe Institute of Technology (KIT), D-76021 Karlsruhe, Germany}
\affiliation{Department of Chemistry, University of Basel, St. Johannsring 19, CH-4056 Basel, Switzerland}
\affiliation{Lehn Institute of Functional Materials (LIFM), Sun Yat-Sen University (SYSU), Xingang Rd. W., Guangzhou, China}
\author{Wulf Wulfhekel}
\affiliation{Institute for Quantum Materials and Technology, Karlsruhe Institute of Technology, D-76021 Karlsruhe, Germany}
\author{Richard Korytár}
\affiliation{Department of Condensed Matter Physics, Faculty of Mathematics and Physics, Charles University, Ke Karlovu 5, 121 16, Praha 2, Czech Republic}

\begin{abstract}
The generation of unidirectional motion has been a long-standing challenge in engineering of molecular motors and, more generally, machines. A molecular motor is characterized by a set of low energy states that differ in their configuration, i.e. position or rotation. In biology and Feringa-type motors, unidirectional motion is driven by excitation of the molecule into a high-energy transitional state followed by a directional relaxation back to a low-energy state. Directionality is created by a steric hindrance for movement along one of the directions on the path from the excited state back to a low energy state.  Here, we showcase a principle mechanism for the generation of unidirectional rotation of a molecule without the need of steric hindrance and transitional excited states. The chemical design of the molecule consisting of a platform, upright axle and chiral rotor moiety enables a rotation mechanism that relies on the transfer of orbital angular momentum from the driving current to the rotor. The transfer is mediated via orbital currents that are carried by helical orbitals in the axle.
%
\end{abstract}

\maketitle


\section{Introduction}
%
Like any other engine, also molecular motors require for operation an energy drive such as light \cite{saywell_light-induced_2016, van_delden_unidirectional_2005}, chemical \cite{leigh_unidirectional_2003, kelly_unidirectional_1999} or electrical energy. For molecular rotors, the most frequently proposed operation principle is that the role of this energy input is to activate the molecule by exciting it from one of the low-energy to a high-energy intermediate state, which subsequently relaxes back to one of the low-energy states. Directionality is gained by the fact that relaxation is asymmetric, i.e. it follows a ratchet potential, and the motor performs one rotational or translational step in the relaxation process. 

Scanning probe microscopy (SPM) with its atomic resolution is typically used to observe motion on the scale of individual molecules. The molecules are mounted on a substrate and their motion under external drive is directly observed \cite{kudernac_electrically_2011, tierney_experimental_2011, simpson_adsorbate_2023}.
In order to investigate this directed rotation with SPM, the molecule needs to be immobilized on a surface and the rotation occurs typically around an axis normal to the substrate. The activated states are usually too short lived to be detected with the relatively slow SPM techniques. Hence, the rotation of the molecule is witnessed by transitions between the relaxed (locked) states. In this sense, the observation is stroboscopic. At least three locked states are required to determine a winding sense of the rotation. 

\begin{figure}[b]
\includegraphics[width=0.75\columnwidth, angle=0]{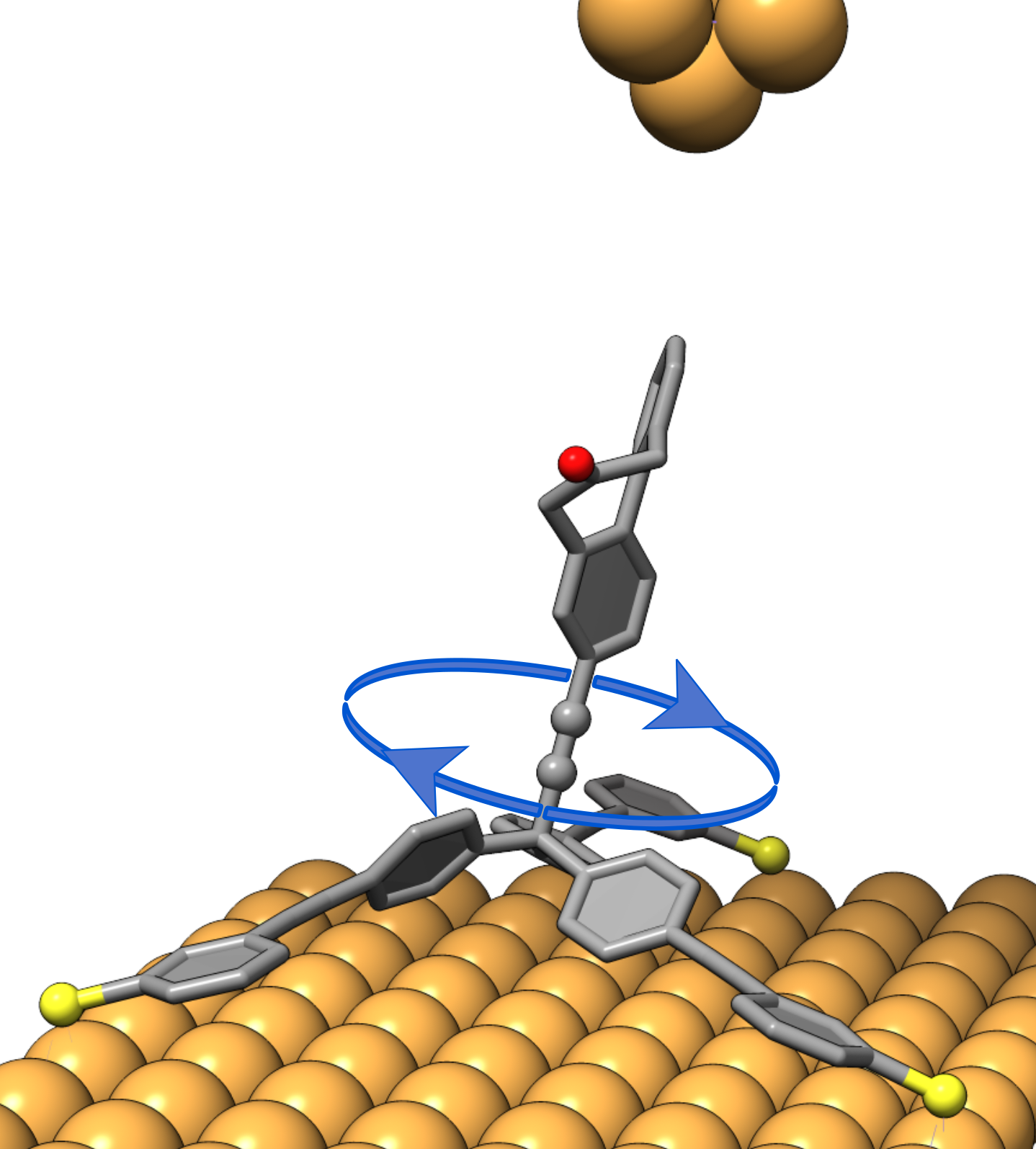}
\caption{Geländer group (rotor) mounted on an organic tripod (stator). The axle connecting both is made by a triple bond that allows for an easy rotation. }
\label{f1}
\end{figure}

The pivotal atom or bond may connect directly to the substrate \cite{tierney_experimental_2011, mishra_current-driven_2015, Stolz2020, bauer_tip-induced_2020}, or may be an integral part of the molecule itself \cite{Perera2013, JasperTonnies2020}. The latter case is particularly appealing as it increases structural and spatial control over a rotor’s subunits, but it is more demanding with respect to molecular design and synthesis. The approach requires a foot structure that immobilizes the molecule but also controls the molecule’s spatial arrangement on the substrate, and a perpendicularly arranged revolving axis on which the rotor is mounted \cite{Valášek2016}. Another advantage of this strategy is the ease of rotation, as the division into the above-mentioned subunits also lifts the rotor, and thereby reduces interaction with the substrate (see Fig. \ref{f1}).

In many scanning tunneling microscopy (STM) experiments reporting a unidirectional rotation, the rotation is driven by current pulses at specific locations within the molecule. To keep the rotation going, the STM-tip must rotate along. The rotation is documented by comparing STM images before and after the pulse \cite{kudernac_electrically_2011, Perera2013}. Directionality in this approach is affected by the cyclic path of the STM tip \cite{Homberg2019, frauhammer_addressing_2021}. 

Experimental protocols have been proposed that operate with a stationary tip and drive the rotor with a dc-current. The different locked positions are then detected by their individual current levels in the time trace of the tunneling current $I(t)$. This method avoids an influence of the tip motion on the molecular rotation at the cost of full imaging of the rotational stages. Instead, the directionality is witnessed by the sequence of the tunneling current levels \cite{tierney_experimental_2011, Homberg2019, Stolz2020, bauer_tip-induced_2020, JasperTonnies2020}.

An interesting open question to be investigated in the context of current-driven molecular motors relates to the microscopic origin of the preferred sense of rotation. Previous explanations invoke a ratchet potential in which the activated state relaxes with a preference in one direction \cite{tierney_experimental_2011, Perera2013, mishra_current-driven_2015, JasperTonnies2020, Stolz2020, bauer_tip-induced_2020}. 

In this work, a molecular motor has been designed and synthesized that contains a rotor in the form of a chiral "Geländer" group which is mounted on a rotatable carbon triple bond on top of a tripodal foot structure shown in Fig. \ref{f1}. 
Our theoretical investigation based on density functional theory (DFT) transport calculations suggest that the rotation of this rotor 
is driven by a mechanism that has not been discussed before, i.e., the orbital momentum transfer mediated via circulating currents flowing through helical orbitals - here realized along the triple bond. Moreover, the DFT study shows that the frontier orbitals of the triple bond occur in pairs of opposite helicity. It turns out that this generic observation implies the peculiar prediction that the motor turns the same way, irrespective of the current direction, when only one helical orbital is involved in charge transport. 
Experimentally, we find a high directionality of rotation associated with transport through the helical orbitals of the triple bond, which is consistent with the theoretical predictions.

\section{Theory}

Currents flowing through chiral molecular structures, such as the Gel\"ander molecule shown in Fig. \ref{f1}, may carry an orbital angular momentum that may exert a mechanical torque on the molecule, and thus force it to rotate. 
%
This mechanism is essentially classical and has been explored recently within a minimal model \cite{Korytar2023}. In this paper we present a quantum-mechanical analogue of this phenomenon; it exploits the fact that individual molecular orbitals can be helical and thus carry (transverse) circulating currents, even though the supporting atomic structure is not helical. 


Our analysis for the molecule of the type shown in Fig. \ref{f1} employs density functional theory with PBE approximation~\cite{pbeFunctional} as implemented
in Ref.~\cite{TURBOMOLEVERSION,TURBOMOLE,AlrichsBasis}. 
The transport simulations employ the non-equilibrium Green's function (NEGF) formalism \cite{LipavskyVelickyGKB,LipavksyTextbook,aitranss,CuevasTextbook,BruusTextbook} based on Kohn-Sham energies and orbitals as implemented in Ref.~\cite{Walz2015,WalzThesis}, which also provides the spectrally resolved current densities, 
$\mathbf{j}(\mathbf{r}, E)$.
Associated with this current density is an orbital angular momentum, in particular a component $L_z(E)$ along the axis of the Gel\"ander molecule; see Supplementary Information for details. 

\begin{figure}[b]
\includegraphics[width=\columnwidth]{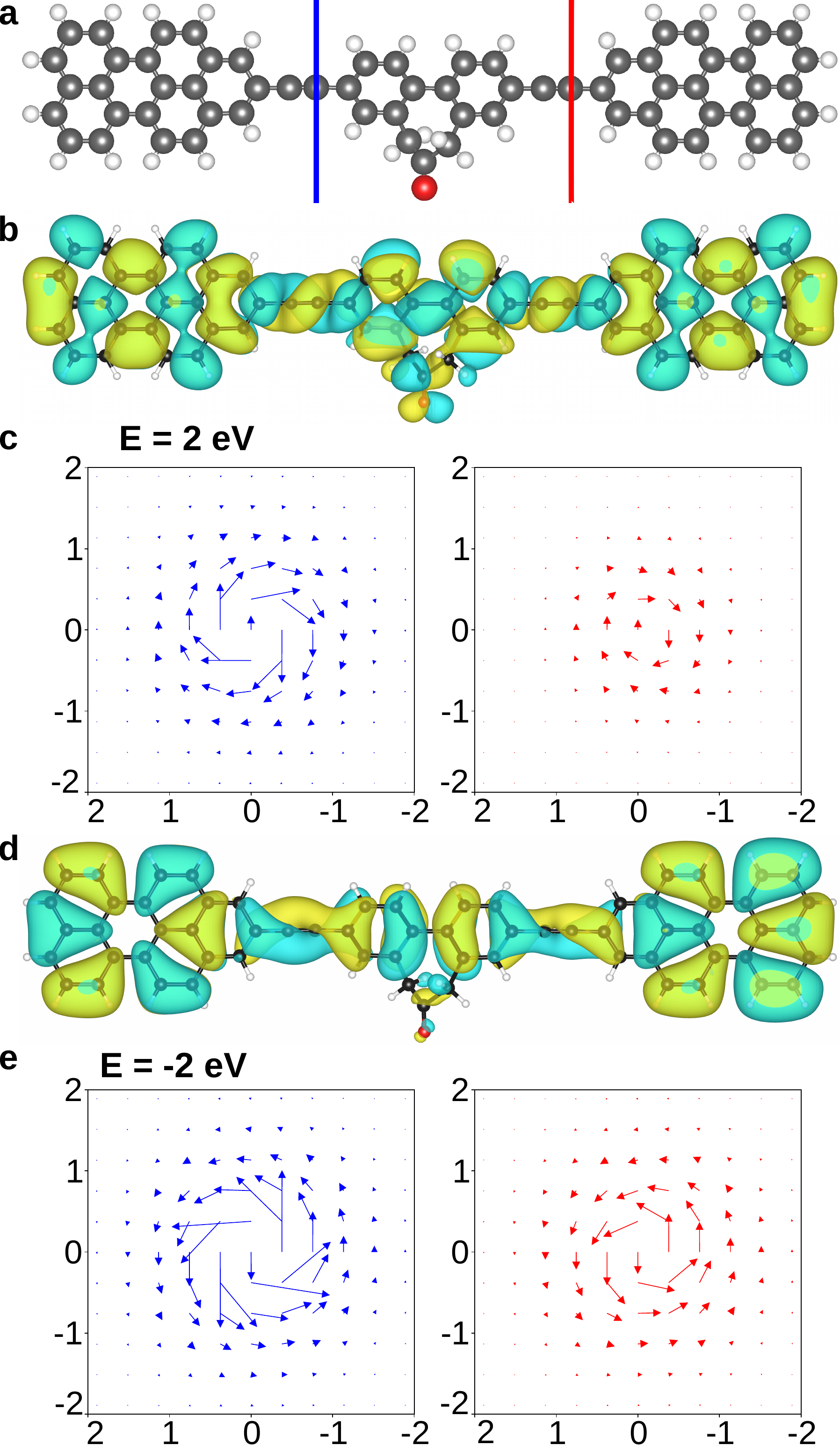}
\caption{\textbf{Circulating currents in helical molecular orbitals.} \textbf{a} Symmetrized atomic structure replicating the functional parts of the Geländer molecule shown in \figref{f1}. \textbf{b} Wavefunction isosurface plot of transport orbital above Fermi energy, displaying helical features, and \textbf{c} spectrally resolved current density at 2~eV above the Fermi energy projected onto cross-sectional plane along the triple bonds (blue and red lines). \textbf{d} Wavefunction isosurface plot of transport orbital below Fermi energy, displaying different helical features, and \textbf{e} analogue to \textbf{c} for the orbital \textbf{d}.}
\label{f2}
\end{figure}

To keep the calculations tractable, we focus on the central element of the device, Fig.~\ref{f1}, i.e. the Geländer group in conjunction with the connecting triple bond, which acts as a molecular pivot. Replacing the full system -- Geländer molecule plus tripod -- by an effective (symmetrized) variant results in the atomistic structure shown in Fig. \ref{f2}a. We investigated the influence of the electrode size, size of basis and choice of the numerical integration scheme on the resulting angular momentum predictions 
and conclude that our calculations are converged for qualitative analysis of the angular momentum energy dependence (see Supplementary Information).

The Kohn-Sham wavefunctions representing the occupied helical and unoccupied helical states closest to the Fermi energy are displayed in Figs. \ref{f2}b,d. As one would expect, the orbitals exhibit a transparent nodal structure that for the most part derives from the $\pi-$orbitals of the $sp^2$-hybridized carbon atoms. These orbitals carry the main current flowing between the two electrodes at respective voltages. 
A characteristic feature of these orbitals 
is the helical shape of wavefunctions wrapping around the pivot, i.e. the triple bond. Its origin can be traced back to the properties of the isolated triple bond chains: 
$[n]$-cumulenes~\cite{Garner2019,JorgensenSolomonHelical} and carbynes~\cite{GunasekaranHelical}. 
Essentially, the two degenerate molecular orbitals of the formally rotationally invariant triple bonds are remixed by the interaction with the chiral Geländer moiety, and thus turn into the helical segments seen in Fig. \ref{f2}b,d of the wavefunctions of the entire moiety.

The important observation is that these helical segments carry the dominant part of the current-induced orbital angular momentum, since the terminal benzoperylenes are flat, while the central bridged biphenyl Gel\"ander structure triggeres the helical chirality without having helically arranged frontier orbitals itself. 
Notice further that the winding sense of 
the orbital pair, Figs. \ref{f2}b,d, is opposite. 
%

The chirality of the orbitals has direct impact on the current passing through the molecule. \figref{f2}c,e illustrates the current flow ${\bf j}({\bf r},E)$ within a cross-section through the molecule, specifically in the $x{-}y$-plane perpendicular to the triple bond. The curl seen there reflects the circulating current wrapping around the triple bonds seen in \figref{f2}b,d. 
Thus, when electrons enter the chiral orbitals of a triple bond from a flat electrode, they gain an orbital angular momentum that they again lose, when exiting the triple bond. Due to conservation of angular momentum, for this short period, the central part of the molecule, i.e. the rotor, must take up the orbital momentum intermediately and rotate. This aspect of current-induced torque is fully analogous to the classical situation explored in Ref.~\cite{Korytar2023}. In particular, the sign of the resulting torque follows the sign of the orbital angular momentum, as one would expect.  
\begin{figure}[th]
\includegraphics[width=\columnwidth, angle=0]{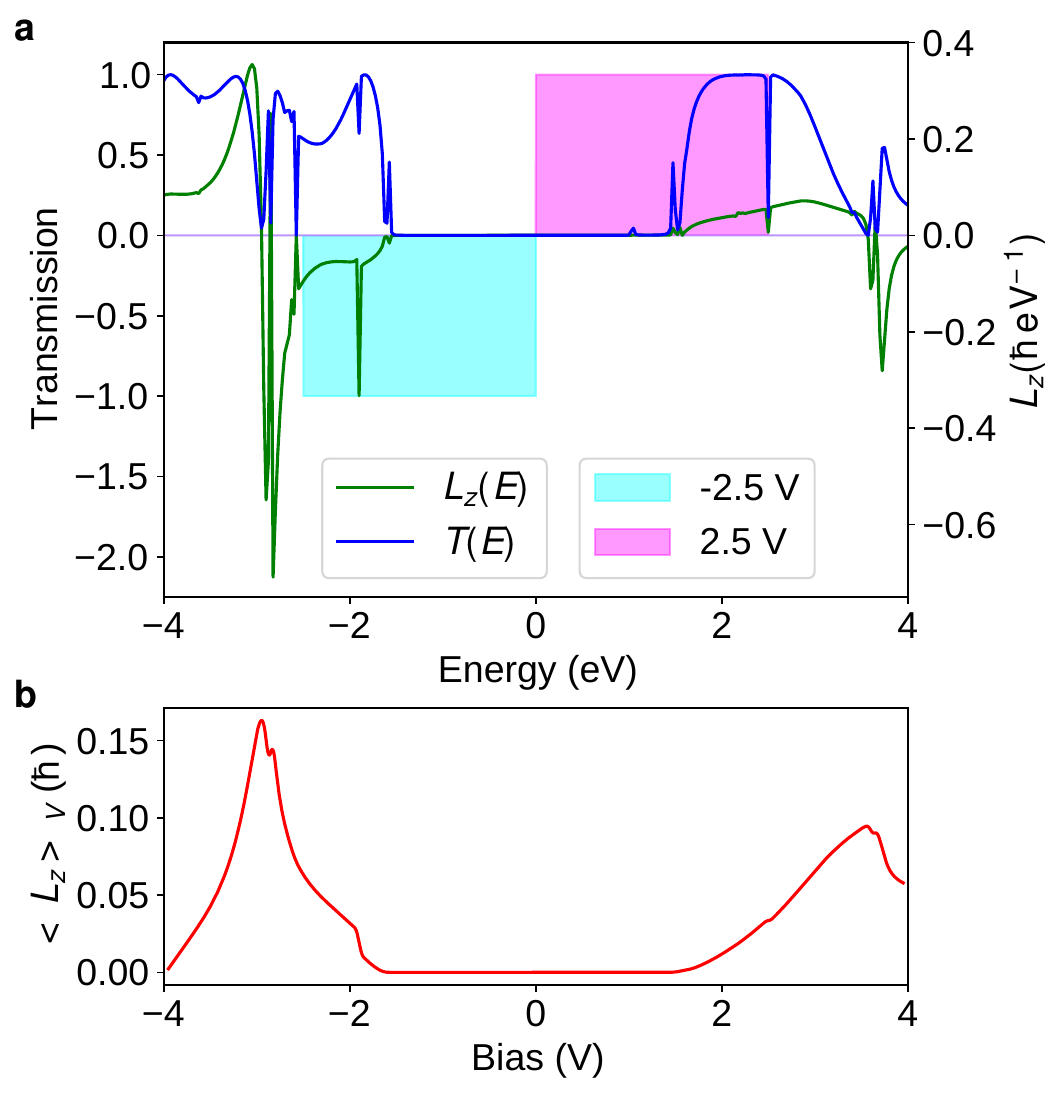}
\caption{ \textbf{Orbital angular momentum and transmission.}
\textbf{a} Current-carried orbital angular momentum $L_z(E)$ (green); also shown is the transmission function $T(E)$ (blue). 
The sign-change in $L_z(E)$ near $E_\text{F}{\equiv}0$ reflects the redirection of the winding sense between the occupied and unoccupied orbitals illustrated in \figref{f2}. 
The filled boxes represent the voltage windows at 
$\Vbias{=}\pm 2.5$~V assuming that the potential of the substrate is fixed.
\textbf{b} The red curve shows the energy integral of $L_z(E)$ with the voltage window for a given bias.
\label{f4}}
\end{figure}

In this spirit, we present in 
Fig. \ref{f4}a the current-induced  orbital angular momentum $L_z(E)$. Indeed, the opposite helicity of orbital pairs reflects in $L_z(E)$: the sign of the dominating parts of $L_z(E)$ for $E<E_\text{F}$ and $E>E_\text{F}$ is opposite. Further, for energies between the two orbitals (Fig.~\ref{f2}b,d), no net orbital angular momentum is observed and the corresponding transmission is small (see dark blue line in Fig. \ref{f4}a).
Thus upon reversing the sign of the bias voltage, the helicity of the current carrying orbitals change together with the current direction and thus the sense of rotation of the motor remains the same.



\begin{figure*}[t]
    \centering
    \includegraphics[width=0.713\linewidth]{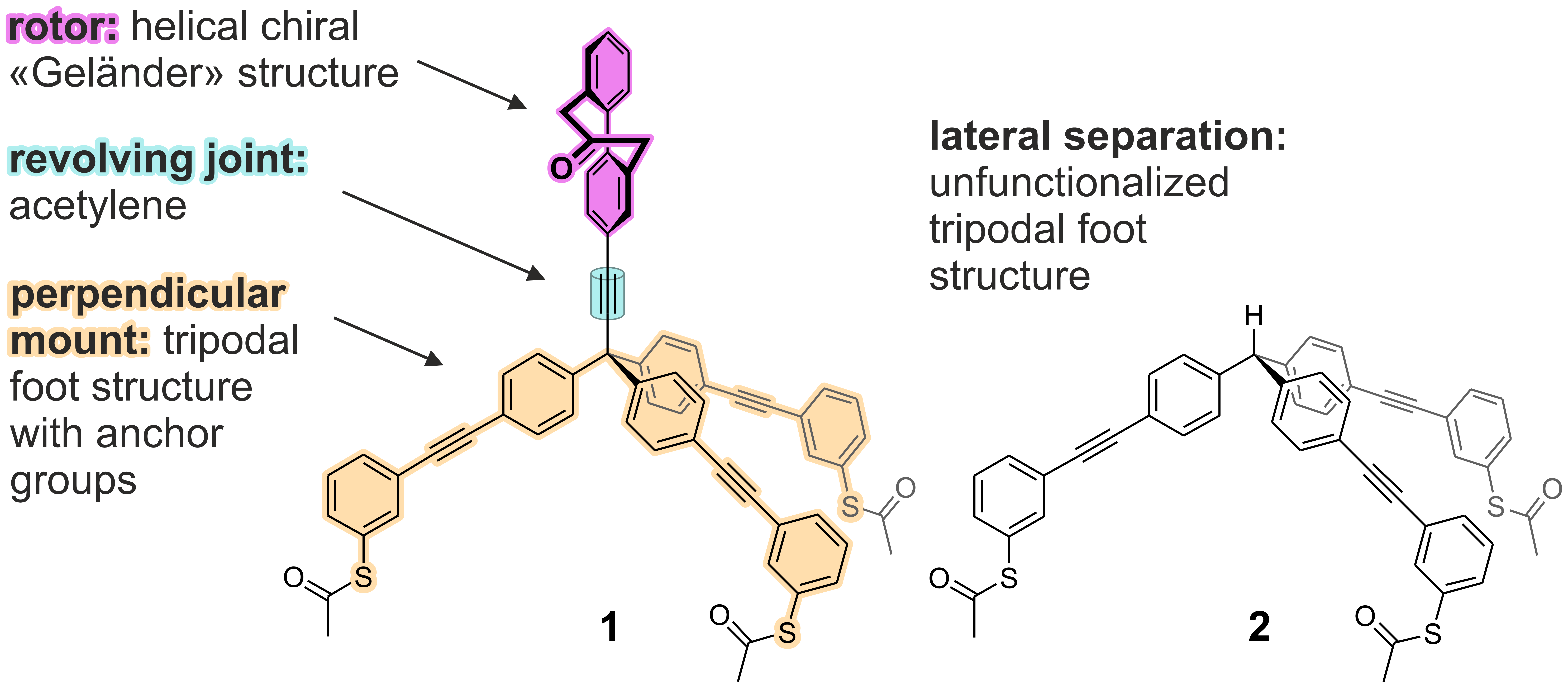}
    \caption{\textbf{The model compounds designed and synthesized for the experiment:} the molecular rotor \textbf{1} divided in its functional subunits and the bare tripodal foot structure \textbf{2} to dilute the density of rotors on the substrate.}
    \label{Molecules}
\end{figure*}


We mention that at energies further away from $E_\text{F}$, $E{\approx} \pm
3.5$eV, a trace of a second pair of helical orbitals is seen in Fig. \ref{f4}a.
Since these orbitals have helicity opposite to the first pair, the associated
orbital angular momentum $L_z(E)$ reverses sign. In an experiment with a bias
voltage $U$, all states in the energy window between $E_F$ and $eU$ contribute
to transport. Therefore, we calculated the energy integral of $L_z(E)$ up to a
given voltage (see Fig. \ref{f4}b). It shows that the net current-induced
mechanical torque is predicted to have local maxima at positive and negative
voltages: if the voltage window becomes large enough, it crosses a root of
$L_z(E)$ and the torque begins to decrease. 

\section{Experiment}

\begin{figure*}[t]
	\centering
	\includegraphics[width= 0.8\textwidth]{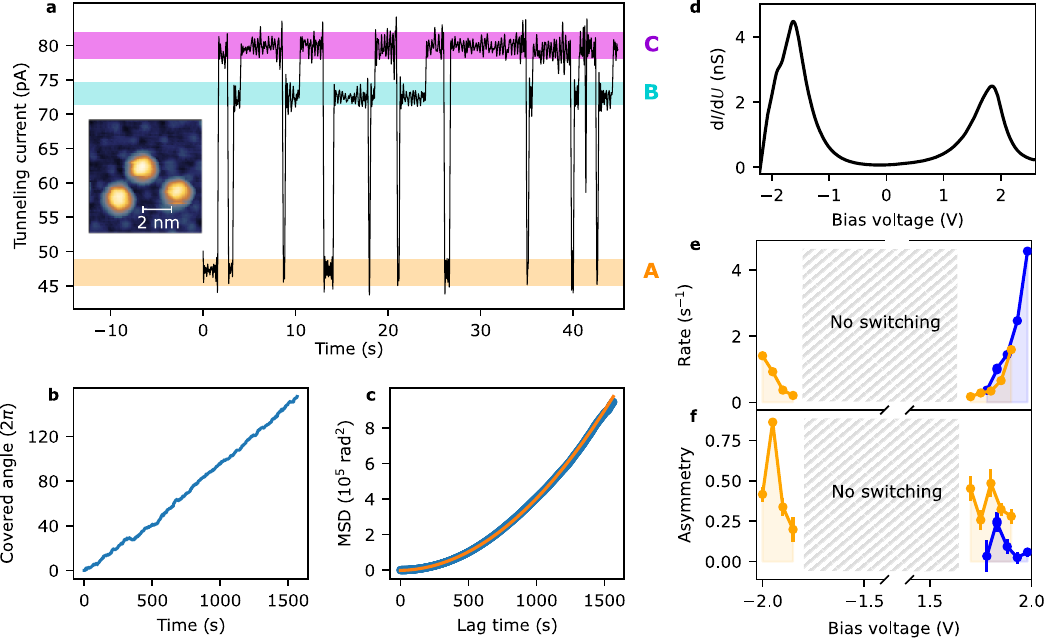}
	\caption{\textbf{Unidirectional rotation in the STM junction.} \textbf{a} Typical time trace of the tunneling current with the STM tip placed above a rotor recorded at a sample voltage of \textit{U} = 1.8 V. The inset shows three rotor head groups protruding from an island of self-assembled tripodal foot structures (recorded at 1.5 V, 30 pA). In the current time trace, three different current levels labeled A,B,C can be distinguished which can be assigned to three different rotational states of the molecular head group. \textbf{b} Covered rotation angle in multiples of 2$\pi$. \textbf{c} Mean square displacement derived from the measurement in \textbf{b} (blue dots) and fit according to $\text{MSD}(\tau) = \left(\omega \tau\right)^2 + 2D\tau + \text{MSD}_0$, where $\tau$ is the lag time, $\omega$ is the angular frequency, $D$ is the rotational diffusion constant. The fit (orange line) provides $\omega^2$ = 0.39 (rad s$^{-1}$)$^2$, equivalent to a rotational frequency of 0.1 Hz, $D$ = 7.07 (rad$^2$ s$^{-1}$)  and $\text{MSD}_0$ = -1900 rad$^2$, equivalent to a dominant quadratic contribution expected for directed rotation. \textbf{d} Differential conductance $dI/dU$ measured with the tip placed above a rotor head group. \textbf{e} Number of transitions per second and \textbf{f} asymmetry between the two switching directions as a function of the sample voltage. Blue and orange markers each correspond to one fixed position of the STM tip above a molecule.}
	\label{STM_combined}
\end{figure*}

The theoretical predictions are detailed and call for comparison with the experiment. To investigate the hypothesized molecular rotor propelled by the current through the triple bond, the model compound \textbf{1} displayed in \figref{f1} and \figref{Molecules} was developed. The molecular rotor consists of an extended tripodal platform \cite{rai_hot_2023}, a rotational axis in the form of a triple bond, and a compact helical chiral rotor in the form of a Geländer type moiety \cite{rotzler_atropisomerization_2012}.
The tripodal platform guarantees the upright orientation of the rotor after deposition and immobilization on the gold substrate. The triple bond (acetylene) connecting the rotor to the tripodal platform was originally intended exclusively as a rotational axis, and only the insights provided by the theoretical section discussed above revealed the decisive importance of the helical orbitals as propelling subunits. The mounted propan-2-one-1,3-diyl-bridged biphenyl acts as a Geländer-type rotor with an enantiomerization barrier of about 45 kJ mol$^{-1}$. 
While the exposed rotor racemizes quickly in solution at room temperature \cite{rotzler_atropisomerization_2012}, its structure is frozen at the low temperature of the STM experiment such that its chiral nature is maintained in the course of the investigation. 
In contrast to the larger Geländer structures developed in the past \cite{rickhaus_induktion_2014}, its compact structure facilitates both, its upright orientation after deposition and its investigation in the STM experiment. To favour and support the upright orientation of the rotor 
\textbf{1}, the bare tripodal foot structure \textbf{2} lacking the axle and the rotor was provided. The syntheses of both model compounds shown in \figref{Molecules} are provided in the Supplementary Information using procedures published in \cite{McKennon1993,sundar_synthesis_2014,vonlanthen_conformationally_2011,rai_hot_2023,bunck_internal_2012, darcy_organocatalytic_2022}. They are fully characterized by $^1$H- and $^{13}$C-NMR spectroscopy, and mass spectrometry (see Supplementary Information).
The molecules \textbf{1} and \textbf{2} were deposited together in a 1/3 ratio by spraying as dichloromethane solution onto a clean Au(111) surface.
Annealing of the sample at about 100$^\circ$C for 1 hour results not only in deprotected thiol anchor groups forming S-Au bonds, but also in islands consisting of periodic patterns of self-assembled tripodal foot structures, which occasionally expose a rotor (see inset of \figref{STM_combined}a). Only the dilution of the rotor \textbf{1} with the naked foot structure \textbf{2} provided samples with immobilized, laterally separated, and thus non-interacting and upright oriented rotors. With the rotor group exposed to the vacuum being free to rotate around the triple bond, the rotation is internal and not related to changes of the adsorption geometry.

As discussed above, reliable information about a potentially preferred direction of rotation upon current injection can only be inferred from a measurement with a fixed tip position, excluding rotation of the molecules due to scanning of the tip \cite{tierney_experimental_2011, Homberg2019, Stolz2020, bauer_tip-induced_2020, JasperTonnies2020}. 


\figref{STM_combined}a shows a typical time trace of the tunneling current recorded with the STM tip fixed off the rotation axis above the rotor group in order to be able to distinguish the different rotational states. The threefold symmetry of the feet naturally provides a threefold energy landscape for rotation with three local minima. 



Indeed, transitions between the three different current levels can be identified (labeled as A,B,C in \figref{STM_combined}a). 
Single meta-stable states could, however, not be imaged in the topographic measurements because the necessary displacement of the STM tip induced transitions to other rotational states while scanning.
In order to test for a directionality of the switching observed in the current traces (see \figref{STM_combined}a), the two possible sequences A,B,C,A,... and C,B,A,C,.... can be assigned to the two opposite rotation senses. 
For example, in \figref{STM_combined}a, the sequence A,B,C,A,... dominates. The expected superposition with random motion, however, requires statistical analysis, for which we chose a binomial test to scrutinize whether the assumption of a random motion (equal number of  rotations in both directions) can be rejected. Out of 180 evaluated measurements, 74 had a rejection probability of $p_\text{reject}\geq 99.9\%$, clearly excluding the hypothesis of a random switching with high statistical significance and thus providing experimental evidence for the presence of directional motion.
In order to further quantify the rotational switching, we defined a velocity by interpreting each discrete transition between the three states as a rotation of the rotor by $\Delta \varphi = \pm  {120}{^\circ}\hat{=}\pm2\pi/3$. This allows to visualize the trajectory as an accumulated total angle $\varphi$ as a function of time (see \figref{STM_combined}b). 
The angular trajectory clearly shows a linear behavior, in contrast to a pure random (Brownian) motion. The physical quantities that describe a driven motion at the nanoscale (the diffusion constant $D$ and the velocity, here the angular frequency $\omega$) can be extracted from a fit to the mean square displacement as described in \figref{STM_combined}c.
Typical rotational frequencies were found to range from 0.1 to 1 Hz (full rotations per second). The individual residence times in the local energy minima are exponentially distributed. i.e. longer and shorter residence times occur. We have detected these states up to the bandwidth of our current amplifier ($\approx$ 1 kHz) limiting us to test for much higher rotational frequencies.

In order to experimentally verify the theoretical predictions, we studied the transport across the molecule and the induced rotational switching as a function of the applied bias voltage. The differential conductance shown in \figref{STM_combined}d shows resonances at about -1.8 V and +1.9 V, indicating transport via molecular orbitals at the corresponding energies. The rotational switching can be characterized by the rate and its asymmetry (defined as the difference between the transitions in the two directions divided by their sum) and the switching rate as a function of the tunneling parameters. 
At fixed z position and varying the bias voltage, we observe no switching
events at bias voltages -1.7 V $<U<$ +1.8 V  and an increase in the rate with
increasing absolute bias voltage as shown in \figref{STM_combined}e. As the
energetic barrier for rotation is of the order of 10 meV and not eV (see DFT
calculations  \cite{FHI-aims} presented in Suppl. Fig.~\ref{SI_barrier}), the observed onset
bias voltages are not related to the barrier but more likely to transport
through specific orbitals of the molecule in agreement with the differential
conductance presented in \figref{STM_combined}d and the theoretical
expectation. 
The switching asymmetry as a function of the applied voltage (see \figref{STM_combined}e) shows a non-monotonic behavior with local maxima at about 1.83 V (molecule represented by blue dots), at 1.8 V and -1.95 V (molecule represented by orange dots). These maxima coincide with the resonances observed in the differential conductance (see \figref{STM_combined}d). At voltages beyond these maxima, when tunneling into further orbitals becomes more likely, the asymmetry drops again to lower values. Most notably, when the direction of the tunneling current is reversed by applying voltage of opposite polarity, the preferred direction of rotation is not reversed. 
Thus, the experimental observations follow the theoretical predictions, that at low bias, no rotation should occur, rotation senses for transport through the occupied and unoccupied helical orbitals are the same due to the inverted applied biases, and the motion loses directionality at higher bias voltages.
Note that experiments with reversed bias voltage polarity are not straightforward, as the rotation of the molecule is not observed by imaging but from the sequence of three current levels. The determination of rotation sense thus required to first correlate the three states and their current for the two polarities of the voltage (for details see Suppl. Fig. \ref{SI_STM_corr}). Further note, that an absolute sense of rotation cannot be inferred from the current traces alone. In this respect, we cannot claim a specific rotation sense for the molecules but restrict ourselves to changes of the directionality.

The dependence of the rotational switching as a function of the tunneling current is non-trivial as in STM because the variation of the current at constant voltage necessarily requires a modification of the z-position of the STM tip which in turn strongly interferes with the switching (see Suppl. Fig. \ref{SI_STM_deps}).


\section{Conclusions}

We have demonstrated a novel mechanism driving an orbital angular momentum transfer from the charge current flowing though chiral orbitals to a molecular rotor. 
As opposed to previously discussed effects, the working principle does not require excited states and ratchet potentials.  
Its characteristic features are (i) that the sense of rotation of the molecule is independent of the applied bias direction, $\Vbias$, and (ii) that the current-induced torque is a non-monotonous function of $\Vbias$. 
%
%
The proposed mechanism is general and might operate in either its classical or quantum variant. Furthermore, the helical orbital currents might contribute to some of the various experimental manifestations of the chirality induced spin selectivity (CISS). Despite of a significant theoretical effort\cite{Evers2022}, the origin of CISS remains unclear due to the lack of strong spin-orbit interaction in organic molecules. However, the orbital currents from helical orbitals may be externally converted to spin currents in heavy element electrodes via the spin-orbit interaction or the exchange split density of states in a ferromagnet.

\section{Methods}

\paragraph{\bf Computational methods.} The electronic structure of a
symmetrized model of the device is calculated in DFT using PBE
\cite{pbeFunctional} and the def2-TZVP basis \cite{KarlsruheBasis}, as
implemented in TURBOMOLE \cite{TURBOMOLE,TURBOMOLEVERSION} suite. Values of
other computational parameters are presented in the SI. The resulting effective
single particle Hamiltonian is connected to a reservoir constructed by
decimation scheme implemented in the TSaint program \cite{WalzThesis,Walz2015}.
This allows for calculation of lesser Green's function in the non-equilibrium
Green's function framework, as detailed in SI. The spectral density of angular
momentum of the current is then evaluated from the current density as (see SI
equation \eqref{e20})
\begin{align}
    L_z(E) = \int \mathrm d^3 \mathbf{r} 
    \left [ \mathbf{r} \times \mathbf{j}(\mathbf{r}, E)\right]_ z
\end{align}
and the expectation value of the observed angular momentum at specific device bias $V$ is given by
\begin{align}
    \langle L_z\rangle_V = \int \mathrm dE L_z(E) \left [ f_\text{L}(E; V) - f_\text{R}(E;V)\right ]
\end{align}
as detailed in the SI equation (12).

The transmission function is calculated by the standard trace over the device subspace \cite{aitranss}
\begin{align}
    T(E) = \text{Tr} \left [ \bm \Gamma ^ \text{A} _ \text{L} \bm G ^ \text{R} (E) \bm \Gamma ^ \text{A} _ \text{R} \bm G ^ \text{A} (E) \right ]
\end{align}
where $\bm \Gamma ^ A _ \text{L/R}$ is the anti-hermitian part of the advanced self-energy induced by the left/right reservoir, respectively, onto the device, $\bm G ^ \text{A/R} (E)$ is the advanced/retarded Green's function in the device region and the trace runs over the device states.

\paragraph{\textbf{Synthetic strategy.}}
The synthesis of target molecules is described in detail in the SI. Synthesis starts from commercially available diphenic acid, which was reduced to diol, brominated a cyclized to ketone. Subsequent bromination led to a mixture of products where the desired monobrominated product was separated by column chromatography on silica gel. Thereafter, halogen exchange to iodine was carried out via copper-catalysed aromatic Finkelstein reaction. The tripodal stator was prepared by reactions described in previous literature \cite{rai_hot_2023}. The final assembly was realized by Sonogashira reaction and after following transprotection with AgBF$_4$ and acetyl chloride, the target molecule \textbf{1} bearing three thioacetate anchoring groups was obtained in good yield.


\paragraph{\textbf{Materials.}} All starting materials and reagents were purchased
from commercial suppliers (\textit{Alfa Aesar, ABCR} (Karlsruhe, Germany),
\textit{Sigma-Aldrich} (Schnelldorf, Germany), \textit{TCI Chemicals} Europe
(Zwijndrecht, Belgium), \textit{Merck} (Darmstadt, Germany) and used
without further purification. Solvents utilized for crystallization,
chromatography and extraction were used in technical grade. Anhydrous
tetrahydrofuran, acetonitrile and dichloromethane were taken from MBraun
Solvent Purification System equipped with drying columns. Triethylamine
was dried and distilled from CaH\textsubscript{2} under nitrogen
atmosphere. TLC was performed on silica gel 60 F254 plates, spots were
detected by fluorescence quenching under UV light at 254 nm and 366 nm.
Column chromatography was performed on silica gel 60 (particle size
0.040--0.063 mm). Compounds \textbf{VII}\cite{rai_hot_2023},
\textbf{IX}\cite{vonlanthen_conformationally_2011} and \textbf{X}\cite{rai_hot_2023} were
prepared according to a published procedure.

\paragraph{\textbf{Equipment and measurements:}} All NMR spectra were recorded on a
Bruker Avance III 500 or Bruker Avance NEO 400 spectrometer at 25$\,^\circ$C in
CDCl\textsubscript{3}, CD\textsubscript{2}Cl\textsubscript{2} or
DMSO-\textit{d\textsubscript{6}}. \textsuperscript{1}H NMR (500 MHz, 400
MHz) spectra were referred to the solvent residual proton signal
(CDCl\textsubscript{3}, \textit{$\delta$}\textsubscript{H~}=~7.24 ppm;
CD\textsubscript{2}Cl\textsubscript{2}, \textit{$\delta$}\textsubscript{H~}=~5.32
ppm; DMSO-\textit{d\textsubscript{6}}, \textit{$\delta$}\textsubscript{H~}=~2.50
ppm). \textsuperscript{13}C NMR (126 MHz, 101 MHz) spectra with total
decoupling of protons were referred to the solvent
(CDCl\textsubscript{3}, \textit{$\delta$}\textsubscript{C~}=~77.23 ppm;
CD\textsubscript{2}Cl\textsubscript{2},
\textit{$\delta$}\textsubscript{C~}=~54.00 ppm; DMSO-\textit{d\textsubscript{6}},
\textit{$\delta$}\textsubscript{C~}=~39.51 ppm). For correct assignment of both
\textsuperscript{1}H and \textsuperscript{13}C NMR spectra,
\textsuperscript{1}H-\textsuperscript{1}H COSY, \textsuperscript{13}C
DEPT-135, HSQC and HMBC experiments were performed. EI MS spectra were
recorded with a Thermo Trace 1300-ISQ GC/MS instrument (samples were
dissolved in dichloromethane) and \textit{m/z} values are given along with
their relative intensities (\%) at the ionisation voltage of 70 eV.
High-resolution mass spectra were recorded with a Bruker Daltonics (ESI
microTOF-QII) mass spectrometer. IR spectra were recorded with a Nicolet
iS50 FTIR spectrometer under ATP mode. Analytical samples were dried at
40--100$\,^\circ$C under reduced pressure ($\approx$ 10\textsuperscript{-2}
mbar). Melting points were measured with a Büchi Melting point M-560
apparatus and are uncorrected. Elemental analyses were obtained with a
Vario MicroCube CHNS analyser. The values are expressed in mass
percentage. The NMR spectra of all new molecules are shown in the Supplementary Information along
with the molecular structure and atom numbering used for the full assignments of signals in the spectra.

\paragraph{\bf STM experiments.} 
Molecular motor \textbf{1} comprising a platform, a triple bond axis and a Geländer group was deposited in a 1/3 ratio with the bare tripodal structure \textbf{2} onto a Au(111) surface cleaned by repeated cycles of sputtering and annealing. 
Therefore, the clean Au(111) crystal is placed directly opposite to a pinhole in a rough vacuum chamber ( $\approx 1\times10^{-2}$ mbar). About 1 ${\mu}$L of molecular solution (1 mg of molecules in 1 mL of dichloromethane) is being sucked in through the pinhole so that small droplets of the solution land on the sample, similar to our previous works \cite{valasek_synthesis_2014}. Then the sample is transferred into the UHV chamber and annealed at $\approx$ 373 K to promote the on-surface cleavage of the acetyl protection groups and to facilitate the correct orientation of molecules with the sulfur anchors attached on the Au(111) surface. Together with the bare platforms, ordered islands are formed which further stabilizes the upright orientation of the rotors and allows for isolated exposed rotors, thereby minimizing lateral interactions \cite{frauhammer_addressing_2021}. After annealing, the sample was transferred to our custom-built low temperature UHV STM. 

Movement of the rotor is derived from time traces of the tunneling current at a fixed position of the tip. The signal is filtered using FFT and the sequence of different rotational states is automatically detected using a Schmitt trigger. Statistical significance for a directed rotation has been tested using a binomial test with threshold for rejection of the assumption of random motion greater than 99.9$\% $.
\vspace{.5cm}

\paragraph{\bf Acknowledgements.}
Š. Marek acknowledges the funding from Charles University through GAUK project number 366222. The computational resources were provided by the Ministry of Education, Youth and Sports of the Czech Republic through the e-INFRA CZ (ID:90254). R. Korytár acknowledges the support of Czech Science foundation (project no. 22-22419S).
J.S., N.B., and J.W.~acknowledge financial support by the DFG Grants GE 2989/2, MA 2605/6-1 and WI5664/3-1, respectively.
J.W.~and F.E.~acknowledge support from the DFG through SFB~1277 (project no.~314695032, subproject A03) and GRK~2905 (project no.~502572516). 
M.C.-G. acknowledges support from the Gobierno Vasco-UPV/EHU Project No. IT1569-22.

This paper is dedicated to the memory of Josef Michl, an extraordinary scientist, beloved mentor and inspiration to many.

\paragraph{\bf Authors contribution.}
J.S., S.M. and N.B. contributed equally to this work.
W.W., M.M. and F.E. conceived the study. J.L., N.B., M.V. and M.M. designed and synthesized the molecules. J.S., L.G. and W.W. carried out STM experiments and performed analysis of the experimental data. S.M., J.W., F.E., M.C.-G. and R.K. developed the theoretical model and performed DFT calculations. S.M., J.W., M.V., L.G. F.E. M.M. W.W. and R.K. co-wrote the paper. Figures were prepared by J.S., S.M., J.W., M.V., L.G., F.E., M.M. and R.K.. All authors commented on the manuscript.

\paragraph{\bf Data availability.}
The data that support the findings of this study are available upon reasonable request from the corresponding authors.

\paragraph{\bf Code availability.}
The data that support the theoretical findings were acquired using a custom simulation software TSaint \cite{Walz2015, WalzThesis}.
STM data were acquired and analyzed using a custom code. The custom software is available from the corresponding
authors upon reasonable request.

\section*{Competing interests}
The authors declare no competing interests.

\section*{References}
\bibliography{short_refs.bib}

\end{document}


\makeatletter
\renewcommand{\fnum@figure}{\textbf{Suppl. Fig.~\thefigure}}

\newcommand{\ci}{{\mathfrak i}}

\title{Supplementary Information: An electrical molecular motor driven by angular momentum transfer}

\author{Julian Skolaut}
\affiliation{Institute for Quantum Materials and Technology, Karlsruhe Institute of Technology, D-76021 Karlsruhe, Germany}
\author{Štěpán Marek}
\affiliation{Department of Condensed Matter Physics, Faculty of Mathematics and Physics, Charles University, Ke Karlovu 5, 121 16, Praha 2, Czech Republic}
\author{Nico Balzer}
\affiliation{Institute of Nanotechnology, Karlsruhe Institute of Technology (KIT), D-76021 Karlsruhe, Germany}
\author{María Camarasa-Gómez}
\affiliation{Departamento de Polímeros y Materiales Avanzados: Física, Química y Tecnología, Facultad de Química, UPV/EHU,
 20018 Donostia-San Sebastián, Spain}
\affiliation{Centro de Física de Materiales CFM/MPC (CSIC-UPV/EHU), 20018 Donostia-San Sebastián, Spain}
\author{Jan Wilhelm}
\affiliation{Institute  of  Theoretical  Physics and Regensburg Center for Ultrafast Nanoscopy (RUN),  University  of  Regensburg,   D-93053  Regensburg,  Germany}
\author{Jan Lukášek}
\affiliation{Institute of Nanotechnology, Karlsruhe Institute of Technology (KIT), D-76021 Karlsruhe, Germany}
\author{Michal Valášek}
\affiliation{Institute of Nanotechnology, Karlsruhe Institute of Technology (KIT), D-76021 Karlsruhe, Germany}
\author{Lukas Gerhard}
\affiliation{Institute for Quantum Materials and Technology, Karlsruhe Institute of Technology, D-76021 Karlsruhe, Germany}
\author{Ferdinand Evers}
\affiliation{Institute  of  Theoretical  Physics and Regensburg Center for Ultrafast Nanoscopy (RUN),  University  of  Regensburg,   D-93053  Regensburg,  Germany}
\author{Marcel Mayor}
\affiliation{Institute of Nanotechnology, Karlsruhe Institute of Technology (KIT), D-76021 Karlsruhe, Germany}
\affiliation{Department of Chemistry, University of Basel, St. Johannsring 19, CH-4056 Basel, Switzerland}
\affiliation{Lehn Institute of Functional Materials (LIFM), Sun Yat-Sen University (SYSU), Xingang Rd. W., Guangzhou, China}
\author{Wulf Wulfhekel}
\affiliation{Institute for Quantum Materials and Technology, Karlsruhe Institute of Technology, D-76021 Karlsruhe, Germany}
\author{Richard Korytár}
\affiliation{Department of Condensed Matter Physics, Faculty of Mathematics and Physics, Charles University, Ke Karlovu 5, 121 16, Praha 2, Czech Republic}

\maketitle

\tableofcontents

\clearpage

\section{Supplementary theory}
\subsection{Current induced angular momentum density
\label{angularMomentumSection}}
We are interested in the angular momentum that is associated with the current flow through the device. Without any loss of generality, we can define $z$-axis to be along the (symmetrized) device axis, so only the $z$ component $\ell_z$ of the angular momentum is relevant. The goal here is to derive a formula for $\ell_z$ within the conventional framework of the non-equilibrium Green's function approach. \cite{BruusTextbook,CuevasTextbook,aitranss}

\subsubsection{Recapitulation: single-particle operators}
Within a one-particle theory the (orbital) angular momentum operator is given as
\begin{align}
    \hat\ell_z = \hat x \hat p _ y - \hat y \hat p _ x \label{angularMomentumEq}
\end{align}
Its relation to the current density, 
\begin{align}
    \hat j _ x (\mathbf{r}) = \frac{1}{2m} \left \{ \hat p _ x, \ket{\mathbf{r}}\bra{\mathbf{r}} \right \},
\end{align}
reads 
\begin{align}
    \hat\ell_z = m \int d ^ 3 r x \hat j _ y (\mathbf{r}) - y \hat j _ x (\mathbf{r}), \label{e12}
\end{align}
and analogously for the expectation values 
\begin{align}
    \ell_z= m \int d ^ 3 r \left ( \mathbf{r} \times 
        {\mathbf{j}} (\mathbf{r}) \right ) _ z
\end{align}
defined in the one-particle Hilbert space; the definitions $\ell_z\coloneqq \langle \hat \ell_ z \rangle$ and 
${\mathbf{j}}(\mathbf{r})\coloneqq\langle \hat{\mathbf{j}} (\mathbf{r}) \rangle$ have been employed. The corresponding expression in second quantization is given by \cite{BruusTextbook}
\begin{align}
\label{e14}
    \hat L _ z= \sum _ {\mu, \nu} \bra{\mu} \hat \ell_z \ket{\nu} \hat c ^ \dagger _ \mu\hat c _ {\nu}
\end{align}
where $\hat c^\dagger$ and $\hat c$ denote fermionic creation and annihilation operators and $|\nu\rangle, |\mu\rangle$ denote a basis of the one-particle Hilbert space. 

\subsubsection{Non-equilibrium Green's function formalism} 
The expectation value $L_z$ can be obtained from the lesser Green's function~\cite{BruusTextbook}
\begin{align}
    G ^ < _ {\mu \nu} (t - t') = \frac{\ci}{\mathrm{\hbar}} \left \langle \hat c ^ \dagger _ \mu (t) \hat c _ \nu (t') \right \rangle
\end{align}
where the angular brackets denote an average over the many-body Hilbert space and we assumed a stationary state, so that the Green's function is a function of time difference only~\cite{LipavskyVelickyGKB,LipavksyTextbook}. Embarking on \eqref{e14} we have 
\begin{align}
    <L_z> = \sum _ {\mu, \nu} \bra{\mu} \hat\ell_z \ket{\nu} (- \ci \mathrm{\hbar})
    G ^ < _ {\mu \nu} (t = 0). 
\end{align}

Doing the inverse Fourier transform leads to
\begin{align}
    <L _ z>_V = \sum _ {\mu, \nu} \bra{\mu} \hat \ell _ z \ket{\nu}
    (- \ci \mathrm{\hbar} ) \int \frac{dE}{2 \pi \mathrm{\hbar}} G ^ < _ {\mu \nu} (E)
    \label{e17}
\end{align}
and analogous expressions for the $x,y$-directions. We also included the dependence on bias voltage $V$, which enters in the lesser Green's function. Eq. \eqref{e17} 
suggests a spectral decomposition
\begin{align}
    L_i(E) \coloneqq (-\ci \mathrm{\hbar}) \sum _ {\mu, \nu} \bra{\mu} \hat \ell _ i \ket{\nu}G ^ {<'} _ {\mu \nu} (E), \quad i=x,y,z, 
    \label{e18}
\end{align}
where $G^{<} _ {\mu \nu} (E) = G ^ {<'} _ {\mu \nu} (E) (f _ \text{L} (E) - f _ \text{R} (E))$, so that $L_i = \int (dE/2\pi\hbar)\  L_i (E) (f_\text{L} (E) - f_\text{R}(E))$, where $f_\text{L/R}(E)$ is the Fermi distribution function of the left/right reservoir, respectively, which has been factored out of the Green's function for clarity.  
A fully analogous expression also holds for the spectrally resolved contribution to the local current density
\cite{WalzThesis}
\begin{align}
    {\mathbf{j}} (\mathbf{r}, E) &\coloneqq (- \ci \mathrm{\hbar}) \sum _ {\mu, \nu} \bra{\mu} \hat {\mathbf{j}} (\mathbf{r}) \ket{\nu} G ^ {<'} _ {\mu \nu} (E).
    \label{e19}
\end{align}
Combining \eqref{e12}, \eqref{e18} and \eqref{e19}, we obtain 
\begin{align}
{\bf L} (E) = m \int d ^ 3 r \left ( \bm r \times  {\mathbf{j}} (\mathbf{r}, E)\right),
    \label{e20}
\end{align}
and the actual angular momentum is then obtained as integral
\begin{align}
    <L _ z>_V = \int \frac{dE}{2 \pi \mathrm{\hbar}} (f_\text{L} (E) - f _ {\text{R}} (E)) m \int d ^ {3} r (\mathbf{r} \times \mathbf{j} (\mathbf{r}, E)) _ z\,. \label{biasMomentum}
\end{align}

\subsection{Rotation barrier}
\begin{figure}[h!]
\includegraphics[width=0.9\columnwidth, angle=0]{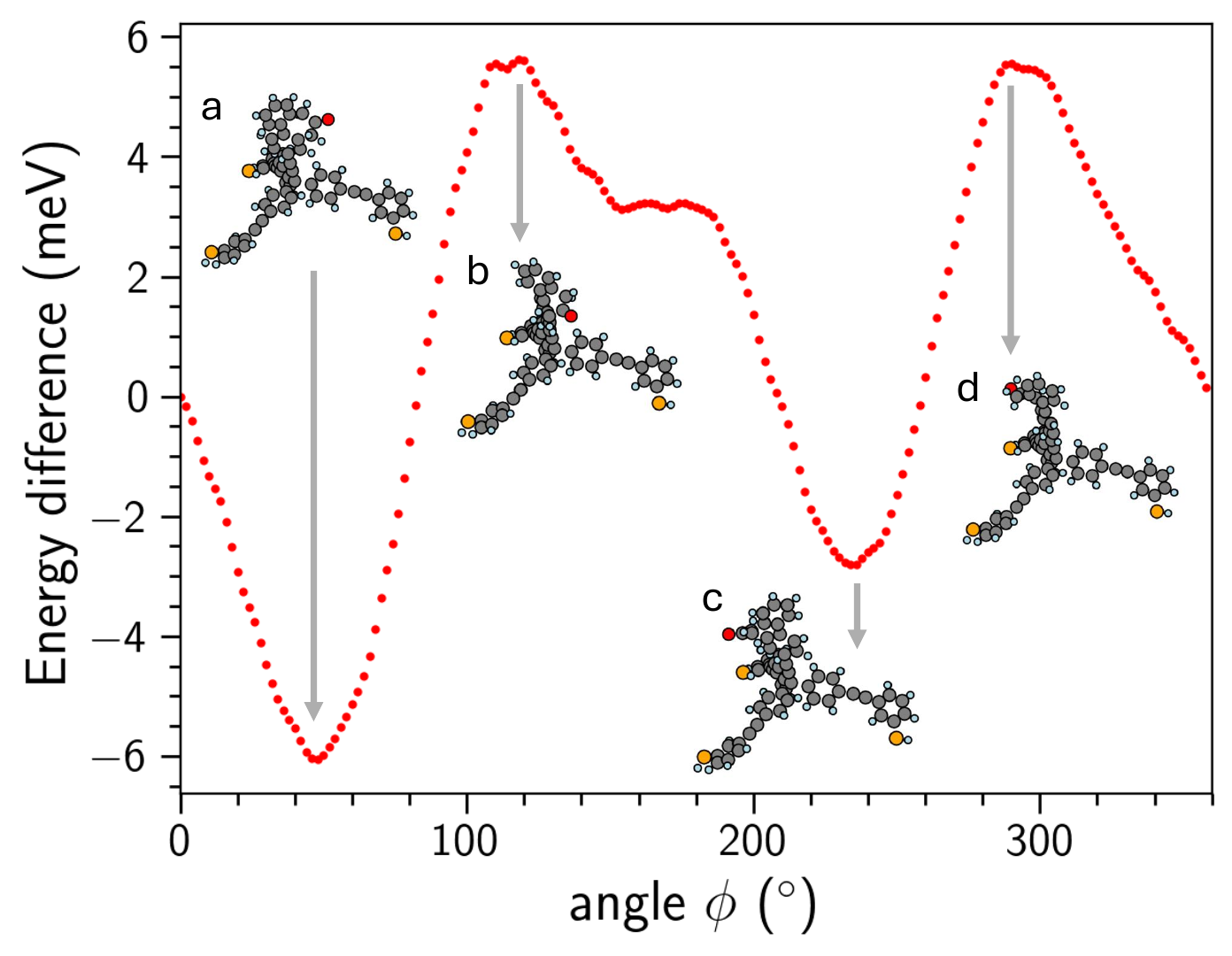}
\caption{Energy change of compound~\textbf{1} as function of the rotation angle of the rotator around the acetylene triple bond (compound geometry in Fig.~\ref{Molecules} in the main text), computed from DFT. 
In the lowest-energy geometry at \textcolor{black}{$\approx$\,48\,$^\circ$} rotation, the oxygen atom points towards a leg of the tripodal platform. \textcolor{black}{Inset: molecule geometries corresponding to the minima (a, c) and maxima (b, d) of the energy barrier as indicated in the plot.} 
}
\label{SI_barrier}
\end{figure}

The rotation of the rotor  by an angle~$\Delta\varphi$ around the triple bond
comes with a change in the total energy of the whole compound~\textbf{1}
(compound geometry in Fig.~\ref{Molecules} in the main text). 
We compute the total energy of compound~\textbf{1} with density functional
theory (DFT) using \textcolor{black}{FHI-aims \cite{FHI-aims} with the}
\textcolor{black}{Perdew-Burke-Ernzerhof (PBE)}
functional~\cite{pbeFunctional}. \textcolor{black}{We employ 'tight` settings,
equivalent to a 'double zeta plus polarization` for the basis sets. We employ
for the ground state total energy calculations default convergence criteria, as
indicated in the FHI-aims package.}   
To obtain the atomic geometry of compound~\textbf{1}, we rotate the rotor by
the angle $\Delta\varphi$ around the triple bond and we relax the three upper
phenyl rings of the tripodal foot structure to allow for atomic rearrangements.
\textcolor{black}{The relaxation is performed using the
Broyden-Fletcher-Shanno-Goldfarb algorithm as implemented in FHI-aims, up to a
threshold value of the force components of 10$^{-2}$ eV/\AA\ per atom.}
We report the change of the total energy  as function of $\Delta\varphi$ in
Suppl. Fig.~\ref{SI_barrier}.
We observe an activation barrier of \textcolor{black}{11.7~meV}, i.e.~the
energy difference between the energetic minimum and maximum. We remark that
the calculated curve is informative about the activation barrier only and not
about the physical potential landscape, which exhibits an actual C$_3$ axis.

\subsection{Computational details}

The electronic structure of the symmetrized device was determined in the TURBOMOLE suite~\cite{TURBOMOLEVERSION,TURBOMOLE} in DFT, using the PBE functional~\cite{pbeFunctional} and def2-TZVP basis sets~\cite{AlrichsBasis}. The atomic positions of the symmetrized device are optimized so that the root mean square of elements of cartesian gradient of energy is less than 0.01~a.u. The electronic density is converged so that the energy change in the self-consistent cycle is less than $10^{-7}$~Hartree. The current density is evaluated in the TSaint code~\cite{WalzThesis}, using the
self-energy of the graphene nanoribbon to couple the device to a reservoir. The self-energy is
determined recursively with a decimation technique~\cite{WalzThesis}.
We have checked the convergence of $L_z(E)$ with respect to the number of self-energy recursions and the basis set size.

\clearpage
\section{Synthesis of target molecules and their characterization}

\subsection{Synthetic approach}
\begin{figure}[h]
\includegraphics[width=0.96\columnwidth]{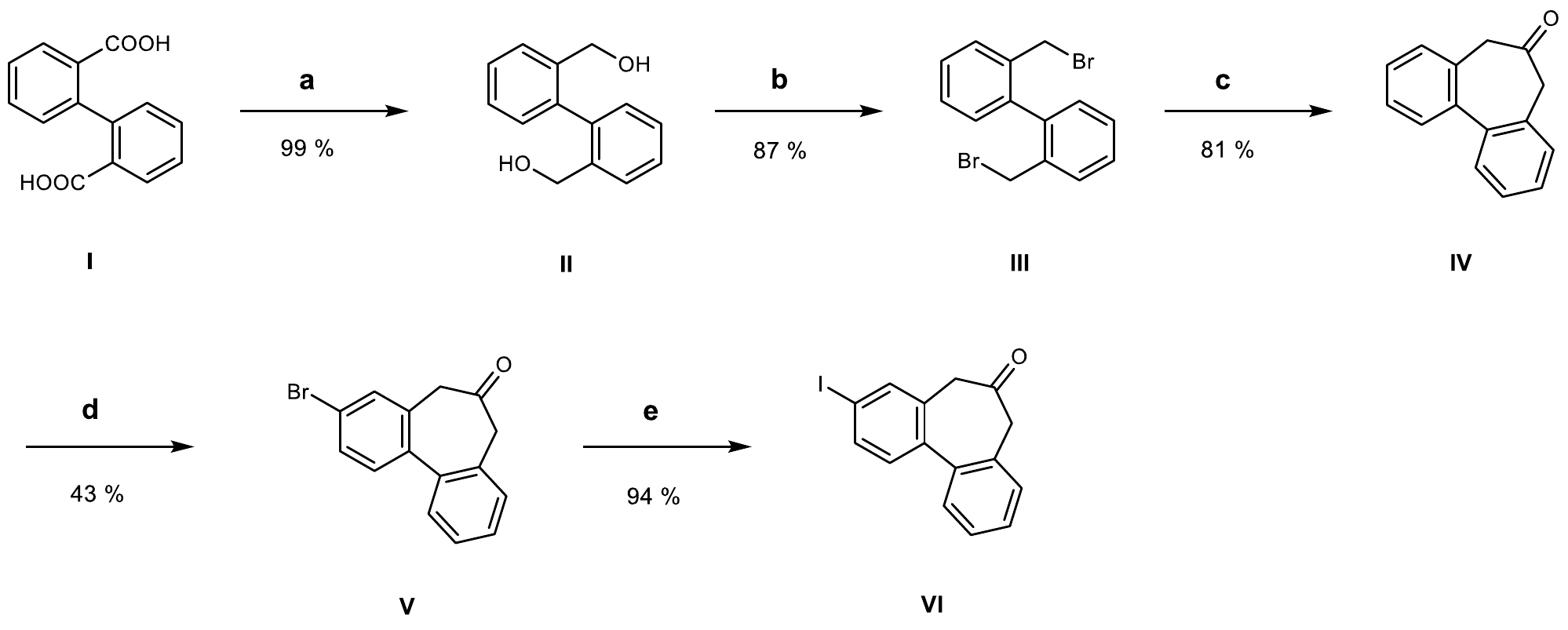}
\caption{\textbf{ Synthetic approach to axially-chiral 2,2'-bridged biphenyl VI.} Reaction conditions: \textbf{a}
NaBH\textsubscript{4}, I\textsubscript{2}, THF; \textbf{b} PPh\textsubscript{3}, Br\textsubscript{2}, CH\textsubscript{3}CN; \textbf{c} TosMIC, TBAB, NaOH, H\textsubscript{2}O, DCM; \textbf{d} AlCl\textsubscript{3}, Br\textsubscript{2}, DCM; \textbf{e} CuI, NaI, \textit{rac}-\textit{N,N'}-dimethylcyclohexane-1,2-diamine, MW, dioxane.}
\label{SyntheticApproach}
\end{figure}

Axially-chiral 2,2'-bridged biphenyl \textbf{VI} with restricted
rotation was prepared by a multistep procedure as described in \sifigref{SyntheticApproach}. The synthesis started from commercially available diphenic acid
\textbf{I}, which was reduced to the corresponding diol \textbf{II} by
the known NaBH\textsubscript{4}/I\textsubscript{2}
system \cite{McKennon1993} in quantitative yield. Subsequent
bromination\cite{sundar_synthesis_2014} by the treatment with
Ph\textsubscript{3}P/Br\textsubscript{2} afforded bromide \textbf{III}
in 87\% yield. The phase transfer reaction using the masked formaldehyde
equivalent TosMIC as a cyclization agent was chosen for the synthesis of
ketone \textbf{IV} in good yield according to a slightly modified
published procedure\cite{vonlanthen_conformationally_2011}. Following bromination of
5,7-dihydro-6\textit{H}-dibenzo{[}\textit{a},\textit{c}{]}{[}7{]}annulen-6-one
\textbf{IV} with Br\textsubscript{2}/AlCl\textsubscript{3} in dry
CH\textsubscript{2}Cl\textsubscript{2} yielded a complex reaction
mixture. According to GC-MS analysis, monobrominated product \textbf{V}
was accompanied by 10\% of the unreacted starting compound, 20\% of the
disubstituted by-product and 10\% of the other two monosubstituted
regioisomers. The pure product \textbf{V} was finally isolated by column
chromatography on silica gel (column length: 1 m, mobile phase:
Hex:CH\textsubscript{2}Cl\textsubscript{2} = 2:1) in 43\% yield. A
cooper-catalysed aromatic Finkelstein reaction was used for halogen
exchange, however, only 50\% conversion was achieved according to GC-MS
analysis. This problem was solved by using microwave heating and the
target compound \textbf{VI} terminated with iodo group was finally
isolated in 93\% yield.

\begin{figure}[h]
\includegraphics[width=0.96\columnwidth]{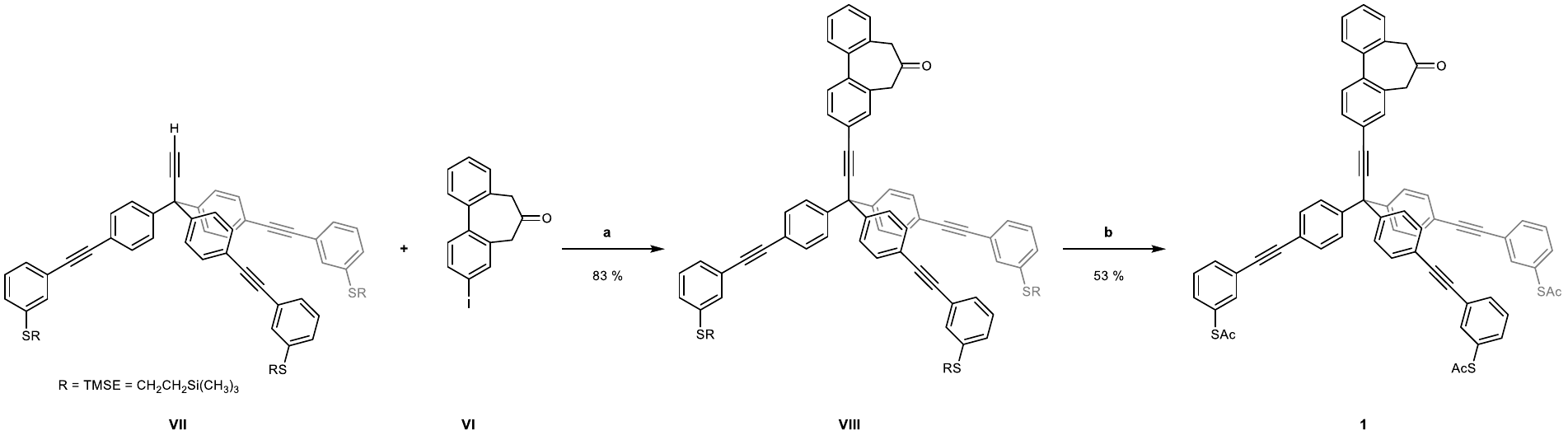}
\caption{ \textbf{Synthesis of tripodal molecular rotor
1.} Reaction conditions: \textbf{a}
Pd(PPh\textsubscript{3})\textsubscript{4}, CuI, Et\textsubscript{3}N;
\textbf{b} AcCl, AgBF\textsubscript{4}, DCM.}
\label{SynthetisOfTripod}
\end{figure}

Final assembly of the tripodal rotor \textbf{1} is outlined in \sifigref{SynthetisOfTripod}, where the iodo-terminated 2,2'-bridged biphenyl \textbf{VI} was first coupled to the extended tripodal platform \textbf{VII} using the Sonogashira cross-coupling protocol to obtain the
2-(trimethylsilyl)ethyl thiol protected (TMSE) tripodal molecule
\textbf{VIII} in 83\% yield. Extended tripodal platform \textbf{VII}
terminated with three thiol anchoring groups was previously prepared
according to a recently published procedure\cite{rai_hot_2023}. It was
found that Sonogashira coupling must be carried out with iodo derivative
\textbf{VI} at low temperature (below 40°C) to prevent
\href{https://www.sciencedirect.com/topics/chemistry/alkyne}{alkyne}
homocoupling, a side reaction competing to
\href{https://www.sciencedirect.com/topics/chemistry/sonogashira-cross-coupling}{Sonogashira
reaction}, which prevails if bromo derivative \textbf{V} is used at
higher temperature instead. Subsequent transprotection of the thiols in
\textbf{VIII} was successfully carried out using AgBF\textsubscript{4}
and acetyl chloride to afford the desired acetyl-protected target
structure \textbf{1} in good yield.

\begin{figure}[h]
\includegraphics[width=0.96\columnwidth]{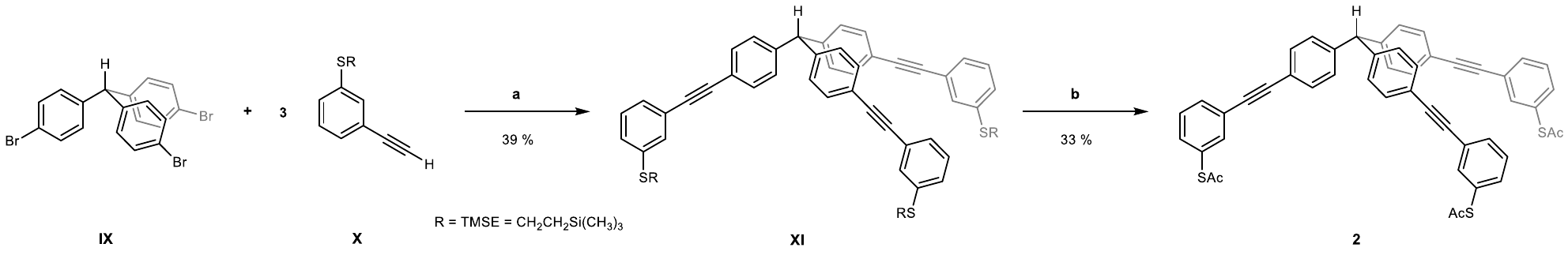}
\caption{ \textbf{Synthesis of dummy tripodal platform
2.} Reaction conditions: \textbf{a}
Pd(PPh\textsubscript{3})\textsubscript{4}, CuI, Et\textsubscript{3}N, 80
°C; \textbf{b} AcCl, AgBF\textsubscript{4}, DCM.}
\label{SynthesisOfDummy}
\end{figure}

Dummy tripodal molecule \textbf{2} was prepared in two reaction steps from the previously prepared tris(4-bromophenyl)methane \textbf{IX}\cite{bunck_internal_2012} and
3-{[}2-(trimethylsilyl)ethylsulfanyl{]}phenylacetylene
\textbf{X}\cite{rai_hot_2023} as shown in \sifigref{SynthesisOfDummy}. First,
tris(4-bromophenyl)methane scaffold \textbf{IX} was coupled with three
equivalents of phenylacetylene derivative \textbf{X} \textit{via}
Sonogashira protocol to provide the extended platform \textbf{XI}.
Subsequent transprotection of the TMSE-protected thiolate \textbf{XI} to
the corresponding thioacetate afforded the desired target molecule
\textbf{2}.

Detailed synthetic procedures for all new compounds and their full
characterization are given bellow. All new compounds were purified by
chromatography and fully characterized by means of conventional NMR and
FTIR spectroscopy, mass spectrometry, as well as elemental analysis.

\subsection{Synthetic Procedures}

\textbf{2,2'-Bis(hydroxymethyl)-1,1'-biphenyl (II)}

Sodium borohydride (7.81 g, 0.21 mmol), diphenic acid \textbf{I} (10 g,
41.3~mmol) and dry THF (200 mL) were added to the oven-dried 1 L
two-necked round bottom flask fitted with a magnetic stirring bar,
reflux condenser and dropping funnel with pressure equalizer. The
reaction mixture was cooled to 0~°C using an ice-water bath. A solution
of iodine (22 g, 86.7 mmol) in 100 mL of dry THF was added dropwise over
60~min and the reaction mixture was heated to reflux for 18 h. Then, 30
mL of methanol was slowly added at room temperature until the mixture
became clear. After 30 min of stirring, the solvent was removed under
reduced pressure and the residue was dissolved in 20\% KOH (200~mL).
This mixture was stirred for 4~h at room temperature before extracted
with EtOAc (3~×~250 mL). Combined organic phase was dried over
MgSO\textsubscript{4}, filtered and all volatiles were evaporated under
reduced pressure. After drying, the product \textbf{II} (9.08 g) was
isolated as a white solid in 99\% yield. Spectroscopic data are
consistent with the literature\cite{sundar_synthesis_2014, darcy_organocatalytic_2022}.
\textsuperscript{1}H NMR (400 MHz, CDCl\textsubscript{3}) $\delta$ (ppm) = 7.47
(dd, \textit{J} = 7.3, 0.8 Hz, 2H, C\textsuperscript{3,3´}H), 7.38 (td,
\textit{J} = 7.5, 1.2 Hz, 2H, C\textsuperscript{4,4´}H), 7.32 (td,
\textit{J} = 7.4, 1.4 Hz, 2H, C\textsuperscript{5,5´}H), 7.13 (dd,
\textit{J} = 7.5, 1.0 Hz, 2H, C\textsuperscript{6,6´}H), 4.33 (m, 4H,
CH\textsubscript{2}), 2.78 (s, 2H, OH); \textsuperscript{13}C NMR (101
MHz, CDCl\textsubscript{3}) $\delta$ (ppm) = 140.2 (C\textsuperscript{1,1´}),
138.8 (C\textsuperscript{2,2´}), 129.9 (C\textsuperscript{6,6´}H), 129.8
(C\textsuperscript{3,3´}H), 128.3 (C\textsuperscript{4,4´}H), 127.9
(C\textsuperscript{5,5´}H), 63.1 (CH\textsubscript{2}).
 
\textbf{2,2'-Bis(bromomethyl)-1,1'-biphenyl (III)}

An oven-dried 250 mL flask was charged with PPh\textsubscript{3} (22.3
g, 85 mmol) and dry CH\textsubscript{3}CN (90 mL). The resulted
suspension was cooled to 0$\,^\circ$C using ice-water bath and bromine (4.4 mL,
85 mmol) was added over 10 min. After that compound \textbf{II} (9.01 g,
42.5 mmol) was added as a solid directly to the stirred reaction mixture
at 0$\,^\circ$C. The stirring was continued for an additional 18 h at room
temperature. The solvent was removed under reduced pressure and the
semi-solid residue portioned between EtOAc (250 mL) and
H\textsubscript{2}O (500 mL), the aqueous phase was extracted with EtOAc
(2~×~150~mL) and combined organic phases were dried over
MgSO\textsubscript{4}. The product was purified by column chromatography
on silica gel (900 g, Hex:CH\textsubscript{2}Cl\textsubscript{2} = 5:1)
to afford 12.6 g of the title compound as a white solid in 87\% yield.
Spectroscopic data are consistent with the
literature\cite{sundar_synthesis_2014, cervantes-reyes_goldi_2019}. R\textsubscript{f} = 0.3 (Hex:CH\textsubscript{2}Cl\textsubscript{2} = 5:1); \textsuperscript{1}H
NMR (500 MHz, CDCl\textsubscript{3}) $\delta$ (ppm) = 7.53 (dd, \textit{J} = 7.7,
1.3 Hz, 2H, C\textsuperscript{3,3´}H), 7.40 (td, \textit{J} = 7.5, 1.5 Hz,
2H, C\textsuperscript{4,4´}H), 7.36 (td, \textit{J} = 7.5, 1.5 Hz, 2H,
C\textsuperscript{5,5´}H), 7.26 (dd, \textit{J} = 7.5, 1.4 Hz, 2H,
C\textsuperscript{6,6´}H), 4.26 (dd, \textit{J} = 76.2, 10.1 Hz, 4H,
CH\textsubscript{2}); \textsuperscript{13}C NMR (126 MHz,
CDCl\textsubscript{3}) $\delta$ (ppm) = 139.6 (C\textsuperscript{1,1´}), 136.1
(C\textsuperscript{2,2´}), 130.9 (C\textsuperscript{3,3´}H), 130.3
(C\textsuperscript{6,6´}H), 128.9 (C\textsuperscript{4,4´}H), 128.5
(C\textsuperscript{5,5´}H), 32.1 (CH\textsubscript{2}).

\textbf{5,7-Dihydro-6\textit{H}-dibenzo{[}\textit{a},\textit{c}{]}{[}7{]}annulen-6-one
(IV)}

Sodium hydroxide (7.56 g, 189 mmol) was dissolved in 72 mL of distilled
water and slowly added to the solution of
2,2´-bis(bromomethyl)-1,1´-biphenyl \textbf{III} (12.6 g, 37.1 mmol),
TosMIC (7.96 g, 40.8 mmol) and tetrabutylammonium bromide (2.87 g, 8.89
mmol) in CH\textsubscript{2}Cl\textsubscript{2} (280 mL) at 0$\,^\circ$C.
Two-phase orange mixture was vigorously stirred at room temperature for
20 h under argon. The organic phase was separated, and the aqueous layer
was extracted with CH\textsubscript{2}Cl\textsubscript{2} (2 × 50 mL).
Concentrated HCl (90 mL) and \textit{tert}-butylmethylether (230 mL) was
added to the combined CH\textsubscript{2}Cl\textsubscript{2} fractions
and the mixture was stirred for 3 h before quenched with
NaHCO\textsubscript{3}. The organic phase was separated and dried over
MgSO\textsubscript{4}, filtered and absorbed on 10 g of silica gel. The
pure product \textbf{IV} (6.27 g) was isolated after column
chromatography on silica gel (900 g) as a colourless oil in 81\% yield.
R\textsubscript{f} = 0.26 (Hex:EtOAc = 10:1); \textsuperscript{1}H NMR
(500 MHz, CDCl\textsubscript{3}) $\delta$ (ppm) = 7.55 (dd, \textit{J} = 7.6, 1.2
Hz, 2H, C\textsuperscript{1,11}H), 7.41 (td, \textit{J} = 7.5, 1.3 Hz, 2H,
C\textsuperscript{2,10}H), 7.34 (td, \textit{J} = 7.5, 1.4 Hz, 2H,
C\textsuperscript{3,9}H), 7.25 (dd, \textit{J} = 7.3, 0.6 Hz, 2H,
C\textsuperscript{4,8}H), 3.55 (dd, \textit{J} = 38.9, 15.5 Hz, 4H,
C\textsuperscript{5,7}H\textsubscript{2}); \textsuperscript{13}C NMR
(126 MHz, CDCl\textsubscript{3}) $\delta$ (ppm) = 210.7 (CO), 139.5
(C\textsuperscript{12,13}), 133.1 (C\textsuperscript{14,15}), 129.6
(C\textsuperscript{4,8}H), 129.5 (C\textsuperscript{1,11}H), 128.3
(C\textsuperscript{3,9}H), 127.9 (C\textsuperscript{2,10}H), 49.5
(CH\textsubscript{2}); MS (EI, 70 eV) \textit{m/z} (\%): 208.1 (86,
{[}M{]}\textsuperscript{+}), 180.1 (75), 179.1 (100), 178.1 (70), 165.0
(88), 152.1 (22), 89.1 (36), 88.1 (11), 76.1 (21), 63.0 (8).

\textbf{3-Bromo-5,7-dihydro-6\textit{H}-dibenzo{[}\textit{a},\textit{c}{]}{[}7{]}annulen-6-one
(V)}

An oven-dried round bottom flask equipped with dropping funnel was
charged with the compound \textbf{IV} (4.06 g, 19.48 mmol) and dry
CH\textsubscript{2}Cl\textsubscript{2} (250 mL) under argon. In a second
flask, bromine (1 mL, 19.52 mmol) was diluted in 125 mL of dry
CH\textsubscript{2}Cl\textsubscript{2} and transferred into the dropping
funnel under argon. Anhydrous AlCl\textsubscript{3} (5.2 g, 40 mmol) was
swiftly added to the flask in one portion and the reaction mixture was
stirred for 20 min at room temperature. Subsequently, a solution of
bromine in CH\textsubscript{2}Cl\textsubscript{2} was slowly added to
the reaction mixture over 2 h. The reaction mixture was quenched by a
saturated aqueous solution of Na\textsubscript{2}SO\textsubscript{3}
(100 mL). The water phase was extracted with
CH\textsubscript{2}Cl\textsubscript{2} (3 × 100 mL), the combined
organic fraction was washed with brine (150 mL), water (200 mL) and
dried over MgSO\textsubscript{4}. The solvent was evaporated under
reduced pressure and the oily residue was absorbed on a small amount of
silica gel. The crude product was purified by column chromatography on
silica gel (550 g, Hex:CH\textsubscript{2}Cl\textsubscript{2} = 2:1).
The pure product \textbf{V} (2.4 g) was isolated as a white solid in
43\% yield. R\textsubscript{f} = 0.28
(Hex:CH\textsubscript{2}Cl\textsubscript{2} = 2:1); m.p. 107.9$\,^\circ$C;
\textsuperscript{1}H NMR (500 MHz, DMSO-\textit{d}\textsubscript{6}) $\delta$
(ppm) = 7.66 (d, \textit{J} = 2.0, 1H, C\textsuperscript{4}H), 7.64 (dd,
\textit{J} = 8.2, 2.1 Hz, 1H, C\textsuperscript{2}H), 7.59 (dd, \textit{J} =
7.6, 1.1 Hz, 1H, C\textsuperscript{11}H), 7.53 (d, \textit{J} = 8.1 Hz,
1H, C\textsuperscript{1}H), 7.46 (td, \textit{J} = 7.4, 1.5 Hz, 1H,
C\textsuperscript{10}H), 7.41 (td, \textit{J} = 7.4, 1.4 Hz, 1H,
C\textsuperscript{9}H), 7.37 (dd, \textit{J} = 7.5, 1.1 Hz, 1H,
C\textsuperscript{8}H), 3.55 (m, 4H,
C\textsuperscript{5,7}H\textsubscript{2}); \textsuperscript{13}C NMR
(126 MHz, DMSO-\textit{d}\textsubscript{6}) $\delta$ (ppm) = 208.7 (CO), 138.1
(C\textsuperscript{13}), 137.7 (C\textsuperscript{12}), 135.4
(C\textsuperscript{14}), 132.8 (C\textsuperscript{15}), 131.9
(C\textsuperscript{4}H), 131.2 (C\textsuperscript{1}H), 130.4
(C\textsuperscript{2}H), 129.5 (C\textsuperscript{8}H), 129.1
(C\textsuperscript{11}H), 128.4 (C\textsuperscript{9}H), 127.8
(C\textsuperscript{10}H), 121.1 (C\textsuperscript{3}), 48.7
(C\textsuperscript{7}H\textsubscript{2}), 47.9
(C\textsuperscript{5}H\textsubscript{2}); FTIR (KBr): $\tilde{\nu}$
(cm\textsuperscript{-1}) = 3059 (w) and 3030 (w, \textit{$\upsilon$}(=CH)), 2926
(m, \textit{$\upsilon$}\textsubscript{as}(CH\textsubscript{2})), 1717 (s,
\textit{$\upsilon$}(C=O)), 1591 (m), 1553 (m, \textit{$\upsilon$}(C=C), Ph), 1475 (s), 1445
(m), 1410 (m), 1390 (m), 1268 (m), 1234 (s), 1152 (w), 1136 (w), 1095
(w), 955 (m), 881 (w), (873 (w), 830 (s), 762 (s), 746 (s); MS (EI, 70
eV) \textit{m/z} (\%): 288.0 (33), 286.0 (33, {[}M{]}\textsuperscript{+}),
260.0 (8), 258.0 (8), 179.1 (100), 178.1 (80), 152.0 (15), 89.1 (32),
88.0 (12), 76.0 (21), 63.0 (6); Elemental anal. calcd. for
C\textsubscript{15}H\textsubscript{11}BrO (285.99): C 62.74, H 3.86;
found: C 63.01, H 3.99.

\textbf{3-Iodo-5,7-dihydro-6\textit{H}-dibenzo{[}\textit{a},\textit{c}{]}{[}7{]}annulen-6-one
(VI)}

An oven-dried argon flushed microwave tube was charged with CuI (20 mg,
0.11 mmol), NaI (313 mg, 2.09 mmol) and compound \textbf{V} (300 mg,
1.05 mmol), followed by addition of anhydrous dioxane (4.5 mL) and
racemic \textit{N,N}'-dimethylcyclohexane-1,2-diamine (33$\,\upmu$L, 0.209 mmol).
Subsequently the septum was exchanged for microwave cup under argon
atmosphere. The reaction mixture was heated in the CEM Discover SP
microwave reactor (MW heating parameters: temperature set to 170$\,^\circ$C,
reaction time 3 h with 1 min of pre-stirring). The target temperature
was reached in $\approx$3 min with the pressure of about 5~bar (dropped to 1 bar
after 50 min), power input during holding time was 60-90 W. After 3 h of
microwave heating, the reaction conversion was almost 99\% according to
GC-MS analysis. The resulting suspension was treated with 25\% aqueous
solution of NH\textsubscript{4}OH (7 mL) and poured into distilled
water. The product was extracted with
CH\textsubscript{2}Cl\textsubscript{2} (3 × 50 mL), dried over
MgSO\textsubscript{4}, filtered and concentrated under reduced pressure.
Column chromatography on silica gel (100 g,
Hex:CH\textsubscript{2}Cl\textsubscript{2} = 1:1) afforded 327 mg of the
desired product in 94\% yield. R\textsubscript{f} = 0.2
(Hex:CH\textsubscript{2}Cl\textsubscript{2} = 1:1); m.p. 96.6~°C;
\textsuperscript{1}H NMR (500 MHz,
CD\textsubscript{2}Cl\textsubscript{2}) $\delta$ (ppm) = 7.77 (dd, \textit{J} =
8.1, 1.8 Hz, 1H, C\textsuperscript{2}H), 7.65 (d, \textit{J} = 1.7 Hz, 1H,
C\textsuperscript{4}H), 7.55 (dd, \textit{J} = 7.6, 1.2 Hz, 1H,
C\textsuperscript{11}H), 7.43 (td, \textit{J} = 7.5, 1.3 Hz, 1H,
C\textsuperscript{10}H), 7.38 (td, \textit{J} = 7.5, 1.4 Hz, 1H,
C\textsuperscript{9}H), 7.32 (d, \textit{J} = 8.1 Hz, 1H,
C\textsuperscript{1}H), 7.28 (dd, \textit{J} = 7.5, 0.6 Hz, 1H,
C\textsuperscript{8}H), 3.40-3.60 (m, 4H,
C\textsuperscript{5,7}H\textsubscript{2}); \textsuperscript{13}C NMR
(126 MHz, CD\textsubscript{2}Cl\textsubscript{2}) $\delta$ (ppm) = 209.3 (CO),
139.6 (C\textsuperscript{13}), 138.9 (C\textsuperscript{12}), 138.5
(C\textsuperscript{4}H), 137.2 (C\textsuperscript{2}H), 135.8
(C\textsuperscript{14}), 133.6 (C\textsuperscript{15}), 131.5
(C\textsuperscript{1}H), 130.0 (C\textsuperscript{8}H), 129.6
(C\textsuperscript{11}H), 128.9 (C\textsuperscript{9}H), 128.3
(C\textsuperscript{10}H), 93.9 (C\textsuperscript{3}), 49.6
(C\textsuperscript{7}H\textsubscript{2}), 49.1
(C\textsuperscript{5}H\textsubscript{2}); FTIR (KBr): $\tilde{\nu}$
(cm\textsuperscript{-1}) = 3058 (w) and 3029 (w, \textit{$\upsilon$}(=CH)), 2925
(w, \textit{$\upsilon$}\textsubscript{as}(CH\textsubscript{2})), 2851 (w,
\textit{$\upsilon$}\textsubscript{sym}(CH\textsubscript{2})), 1716 (s,
\textit{$\upsilon$}(C=O)), 1585 (w), 1546 (w), 1493 (w), 1472 (m), 1446 (w), 1404
(w), 1387 (w), 1267 (w), 1233 (m), 1184 (w), 1153 (w), 1136 (w), 1081
(w), 1002 (w), 995 (w) , 863 (w), 827 (m), 760 (m), 744 (m), 730 (w),
706 (w); MS (EI, 70 eV) \textit{m/z} (\%): 333.96 (84,
{[}M{]}\textsuperscript{+}), 305.9 (13), 179.1 (100), 178.1 (94), 152.0
(19), 151.0 (11), 89.2 (55), 88.0 (17), 76.0 (25), 63.0 (8); Elemental
anal. calcd. for C\textsubscript{15}H\textsubscript{11}IO (333.98): C
53.92, H 3.32; found: C 53.88, H 3.46.

\textbf{Compound (VIII)}

Under inert conditions, in an oven dried Schlenk flask, compound
\textbf{VI} (50~mg, 150$\,\upmu$mol, 1.0 eq),
Pd(PPh\textsubscript{3})\textsubscript{4} (9 mg, 8$\,\upmu$mol, 0.05 eq) and
copper(I)-iodide (3 mg, 15$\,\upmu$mol, 0.1 eq) were dissolved in freshly
distilled triethylamine (1.0 mL) and purged with argon for 30 min. At
40~°C, compound \textbf{VII} (160 mg, 165$\,\upmu$mol, 1.1 eq) dissolved in
freshly distilled and outgassed triethylamine (1.7 mL) was added and the
mixture was stirred for 4 h at this temperature. The reaction mixture
was quenched with NH\textsubscript{4}Cl solution (10\%, 5 mL) and
diluted with ethyl acetate (15 mL). The aqueous phase was extracted with
ethyl acetate (3 × 10 mL) and the combined organic layer was washed with
brine and dried over MgSO\textsubscript{4}. After filtration and
evaporation of all volatiles, the crude product was purified by flash
chromatography on silica gel (100 g, hexane/EtOAc = 40:1,) yielded 147
mg (83\%) of the desired compound as a colourless oil.
R\textsubscript{f} = 0.24 (hexane/EtOAc = 40:1); m.p. 61$\,^\circ$C;
\textsuperscript{1}H NMR (500 MHz, CDCl\textsubscript{3}) $\delta$ (ppm) = 7.52
-- 7.57 (m, 3H, C\textsuperscript{1'\,',11'\,'}H,
C\textsuperscript{2'\,'}H); 7.49 (dd, \textit{J} = 8.6, 1.8 Hz, 6H,
C\textsuperscript{3',5'}H), 7.40 -- 7.44 (m, 5H, C\textsuperscript{2}H,
C\textsuperscript{4'\,'}H, C\textsuperscript{10'\,'}H), 7.35 (td,
\textit{J} = 7.5, 1.3 Hz, 1H, C\textsuperscript{9'\,'}H), 7.28 -- 7.32 (m,
9H, C\textsuperscript{2',6'}H, C\textsuperscript{6}H), 7.21 -- 7.27 (m,
7H, C\textsuperscript{4,5}H, C\textsuperscript{8'\,'}H), 3.53 (m, 4H,
C\textsuperscript{5'\,',7'\,'}H\textsubscript{2}), 2.94 -- 2.97 (m, 6H,
CH\textsubscript{2}-S), 0.90 -- 0.94 (m, 6H, CH\textsubscript{2}-TMS),
0.03 (s, 27H, CH\textsubscript{3}, TMS); \textsuperscript{13}C NMR (126
MHz, CDCl\textsubscript{3}) $\delta$ (ppm) = 209.7 (C\textsuperscript{6'\,'}O),
144.9 (C\textsuperscript{1'}), 139.6 (C\textsuperscript{13'\,'}), 138.8
(C\textsuperscript{12'\,'}), 138.0 (C\textsuperscript{3}), 133.3
(C\textsuperscript{14'\,'}), 133.1 (C\textsuperscript{15'\,'}), 132.7
(C\textsuperscript{4'\,'}H), 131.8 (C\textsuperscript{3',5'}H), 131.5
(C\textsuperscript{2}H), 131.1 (C\textsuperscript{2'\,'}H), 129.7
(C\textsuperscript{8'\,'}H), 129.6 (C\textsuperscript{11'\,'}H), 129.4
(C\textsuperscript{1'\,'}H), 129.3 (C\textsuperscript{2',6'}H), 129.0
(C\textsuperscript{4}H), 128.9 (C\textsuperscript{5}H), 128.9
(C\textsuperscript{6}H), 128.6 (C\textsuperscript{9'\,'}H), 128.0
(C\textsuperscript{10'\,'}H), 123.9 (C\textsuperscript{1}), 122.9
(C\textsuperscript{3'\,'}), 122.3 (C\textsuperscript{4'}), 95.5
(-C\textsuperscript{IV}$\equiv$), 89.7 (-C\textsuperscript{III}$\equiv$), 89.5
($\equiv$C\textsuperscript{II}-), 85.7 ($\equiv$C\textsuperscript{V}-), 56.1
(C\textsuperscript{I}), 49.4
(C\textsuperscript{7'\,'}H\textsubscript{2}), 49.1
(C\textsuperscript{5'\,'}H\textsubscript{2}), 29.5
(CH\textsubscript{2}-S), 16.9 (CH\textsubscript{2}-TMS), -1.5
(CH\textsubscript{3}, TMS); FTIR (ATR\textbf{)}: $\tilde{\nu}$ (cm\textsuperscript{-1}) = 3060 (w), 2950 (w,
\textit{$\upsilon$}\textsubscript{as}(CH\textsubscript{3})), 2919 (w,
\textit{$\upsilon$\textsubscript{a}}\textsubscript{s}(CH\textsubscript{2})), 2895
(w, \textit{$\upsilon$}\textsubscript{sym}(CH\textsubscript{3})), 2852 (w,
\textit{$\upsilon$}\textsubscript{sym}(CH\textsubscript{2})), 2161 (w), 1718 (m,
\textit{$\upsilon$}(C=O)), 1583 (m), 1560 (w), 1502 (m), 1479 (w), 1446 (w), 1401
(w), 1259 (w), 1247 (m), 1161 (w), 1148 (w), 1108 (w), 1094 (w), 1079
(w), 1018 (w), 1007 (w), 888 (w), 854 (m), 824 (m), 781 (m), 764 (w),
750 (m), 729 (w), 684 (m), 578 (w), 560 (w); ESI (+) HRMS calcd for
C\textsubscript{75}H\textsubscript{74}OS\textsubscript{3}Si\textsubscript{3}Na:
1193.4108 {[}M + Na{]}\textsuperscript{+}, found \textit{m/z} 1193.4145.

\textbf{Compound (1)}

In an oven dried Schlenk-flask, compound \textbf{VIII} (77 mg, 66$\,\upmu$mol,
1.0 eq) was dissolved in dry DCM (3.0 mL), cooled down to 0$\,^\circ$C and
flushed with argon. At these conditions, acetyl chloride (0.30 mL) was
added, and the reaction mixture was stirred for 20 min. Subsequently,
AgBF\textsubscript{4} (90 mg, 460$\,\upmu$mol, 7.0 eq) was added. The reaction
mixture was stirred for 3 h to reach the room temperature. After that,
the reaction was quenched with crashed ice (20 g) and diluted with DCM
(5.0 mL), separated, and extracted with DCM (3 × 5 mL). After
evaporation of all solvents at room temperature, the crude product was
purified by flash chromatography on silica (60 g, hexane/DCM = 5:1). The
desired transprotected product \textbf{1} (35~mg, 53\%) was isolated as
a white solid. R\textsubscript{f} = 0.26 (hexane/DCM = 5:1); m.p. 40~°C
(decomposition); \textsuperscript{1}H NMR (500 MHz,
CDCl\textsubscript{3}) $\delta$ (ppm) = 7.58 (bs, 3H, C\textsuperscript{2}H),
7.52 -- 7.56 (m, 6H, C\textsuperscript{6}H,
C\textsuperscript{1'\,',11'\,'}H, C\textsuperscript{2'\,'}H), 7.48 (dd,
\textit{J} = 8.5, 1.7 Hz, 6H, C\textsuperscript{3',5'}H), 7.33 -- 7.45 (m,
9H, C\textsuperscript{4,5}H, C\textsuperscript{4'\,',9'\,',10'\,'}H),
7.31 (dd, \textit{J} = 8.5, 1.6 Hz, 6H, C\textsuperscript{2',6'}H), 7.26
(d, \textit{J} = 7.5 Hz, 1H, C\textsuperscript{8'\,'}H), 3.54 (m, 4H,
C\textsuperscript{5'\,',7'\,'}H\textsubscript{2}), 2.42 (s, 9H,
CH\textsubscript{3}); \textsuperscript{13}C NMR (126 MHz,
CDCl\textsubscript{3}) $\delta$ (ppm) = 209.8 (C\textsuperscript{6'\,'}O),
193.7 (CO), 144.9 (C\textsuperscript{1'}), 139.7
(C\textsuperscript{13'\,'}), 138.9 (C\textsuperscript{12'\,'}), 137.5
(C\textsuperscript{2}H), 134.5 (C\textsuperscript{4}H), 133.3
(C\textsuperscript{14'\,'}), 133.1 (C\textsuperscript{15'\,'}), 132.74
(C\textsuperscript{6}H), 132.72 (C\textsuperscript{4'\,'}H), 131.8
(C\textsuperscript{3',5'}H), 131.1 (C\textsuperscript{2'\,'}H), 129.7
(C\textsuperscript{8'\,'}H), 129.6 (C\textsuperscript{11'\,'}H), 129.4
(C\textsuperscript{1'\,'}H), 129.35 (C\textsuperscript{5}H), 129.32
(C\textsuperscript{2',6'}H), 128.6 (C\textsuperscript{9'\,'}H), 128.5
(C\textsuperscript{3}), 128.1 (C\textsuperscript{10'\,'}H), 124.6
(C\textsuperscript{1}), 122.9 (C\textsuperscript{3'\,'}), 122.2
(C\textsuperscript{4'}), 95.4 (-C\textsuperscript{IV}$\equiv$), 90.2
($\equiv$C\textsuperscript{II}-), 89.1 (-C\textsuperscript{III}$\equiv$), 85.8
($\equiv$C\textsuperscript{V}-), 56.2 (C\textsuperscript{I}), 49.4
(C\textsuperscript{7'\,'}H\textsubscript{2}), 49.1
(C\textsuperscript{5'\,'}H\textsubscript{2}), 30.5
(CH\textsubscript{3}); FTIR (ATR\textbf{)}: $\tilde{\nu}$ (cm\textsuperscript{-~1})
= 2957 (w, \textit{$\upsilon$}\textsubscript{as}(CH\textsubscript{3})), 2921 (w),
2852 (w), 2162 (w), 1713 (w, \textit{$\upsilon$}(C=O)), 1586 (w), 1502 (w), 1465
(w), 1401 (w), 1243 (w) 1108 (w), 824 (w), 787 (w), 766 (w), 752 (w),
684 (w), 612 (w); ESI (+) HRMS calcd for
C\textsubscript{66}H\textsubscript{44}O\textsubscript{4}S\textsubscript{3}Na:
1019.2299 {[}M + Na{]}\textsuperscript{+}, found \textit{m/z} 1019.2307;
Elemental anal. calcd. for
C\textsubscript{66}H\textsubscript{44}O\textsubscript{4}S\textsubscript{3}
(996.24): C 79.49, H 4.45; found: C 79.78, H 4.27.

\textbf{Compound (XI)}

Under inert conditions, in an oven dried Schlenk flask, compound
\textbf{IX} (120~mg, 249$\,\upmu$mol, 1.0 eq),
Pd(PPh\textsubscript{3})\textsubscript{4} (7 mg, 6$\,\upmu$mol, 0.05 eq) and
copper(I)-iodide (2 mg, 12$\,\upmu$mol, 0.1 eq) were dissolved in freshly
distilled triethylamine (2.0 mL) and outgassed with argon for 30 min. At
80~°C, compound \textbf{X} (181 mg, 773$\,\upmu$mol, 3.1 eq) dissolved in
freshly distilled and outgassed triethylamine (2.0 mL) was added and the
mixture was stirred for 16 h at this temperature. Quenching the reaction
with NH\textsubscript{4}Cl (10\%, 20 mL) and diluting with ethyl acetate
(15 mL) was followed by separating of the organic phase and extracting
the aqueous phase with ethyl acetate (3 × 5 mL). The combined organic
layer was washed with brine and dried over MgSO\textsubscript{4}.
Purification by flash chromatography (hexanes/DCM = 9:1) yielded 92 mg
(39\%) of the desired compound as a colourless oil. R\textsubscript{f} =
0.32 (hexanes/DCM = 9:1); \textsuperscript{1}H NMR (500 MHz,
CDCl\textsubscript{3}) $\delta$ (ppm) = 7.46 (d, \textit{J} = 8.1 Hz, 6H,
C\textsuperscript{3',5'}H), 7.42 (s, 3H, C\textsuperscript{2}H), 7.27 --
7.32 (m, 3H, C\textsuperscript{6}H), 7.22 -- 7.26 (m, 6H,
C\textsuperscript{4,5}H), 7.07 (d, \textit{J} = 8.1 Hz, 6H,
C\textsuperscript{2',6'}H), 5.55 (s, 1H, C\textsuperscript{I}H), 2.94 --
2.98 (m, 6H, CH\textsubscript{2}-S), 0.90 -- 0.95 (m, 6H,
CH\textsubscript{2}-TMS), 0.04 (s, 27H, CH\textsubscript{3}, TMS);
\textsuperscript{13}C NMR (126 MHz, CDCl\textsubscript{3}) $\delta$ (ppm) =
143.6 (C\textsuperscript{1'}), 138.0 (C\textsuperscript{3}), 132.0
(C\textsuperscript{3',5'}H), 131.5 (C\textsuperscript{2}H), 129.6
(C\textsuperscript{2',6'}H), 129.0 (C\textsuperscript{4}H), 128.9
(C\textsuperscript{5}H), 128.8 (C\textsuperscript{6}H), 124.1
(C\textsuperscript{1}), 121.7 (C\textsuperscript{4'}), 89.7
($\equiv$C\textsuperscript{II}--), 89.3 (--C\textsuperscript{III}$\equiv$), 56.6
(C\textsuperscript{I}H), 29.5 (CH\textsubscript{2}-S), 16.9
(CH\textsubscript{2}-TMS), -1.5 (CH\textsubscript{3}, TMS); FTIR (ATR):
$\tilde{\nu}$ (cm\textsuperscript{-1}) = 2951 (m,
\textit{$\upsilon$}\textsubscript{as}(CH\textsubscript{3})), 2920 (m,
\textit{$\upsilon$\textsubscript{a}}\textsubscript{s}(CH\textsubscript{2})), 2852
(w, \textit{$\upsilon$}\textsubscript{sym}(CH\textsubscript{2})), 1580 (m), 1559
(m), 1470 (w), 1396 (w), 1260 (m), 1248 (m), 1163 (w), 1010 (w), 883
(w), 836 (m), 780 (m), 751 (m), 728 (w), 682 (m); ESI (+) HRMS calcd for
C\textsubscript{58}H\textsubscript{64}S\textsubscript{3}Si\textsubscript{3}Na:
963.3376 {[}M + Na{]}\textsuperscript{+}, found \textit{m/z} 963.3374.

\textbf{Compound (2)}

In an oven dried Schlenk flask, compound \textbf{XI} (68 mg, 72$\,\upmu$mol,
1.0 eq) was dissolved in dry DCM (2.0 mL), cooled down to 0$\,^\circ$C and
flushed with argon. At these conditions, acetyl chloride (0.2 mL) was
added, and the mixture was stirred for 20 min. Subsequently,
AgBF\textsubscript{4} (98 mg, 505$\,\upmu$mol, 7.0 eq) was added. The reaction
mixture was stirred for 3 h to reach the room temperature. After that,
the reaction was quenched with crashed ice (20 g) and diluted with DCM
(5.0 mL), separated, and extracted with DCM (3 × 5 mL). After
evaporation of all volatiles at room temperature, the crude product was
purified by flash chromatography on silica gel (60 g, hexane/DCM = 1:4).
After drying, the desired transprotected product \textbf{2} (18.3~mg,
33\%) was isolated as a white solid. R\textsubscript{f} = 0.21
(hexane/DCM = 1:4); m.p. 40~°C (decomposition); \textsuperscript{1}H NMR
(500 MHz, CDCl\textsubscript{3}) $\delta$ (ppm) = 7.56 (bs, 3H,
C\textsuperscript{2}H), 7.56 -- 7.52 (dt, \textit{J} = 6.9, 1.8 Hz, 3H,
C\textsuperscript{6}H), 7.45 (d, \textit{J} = 8.3 Hz, 6H,
C\textsuperscript{3',5'}H), 7.34 -- 7.40 (m, 6H,
C\textsuperscript{4,5}H), 7.07 (d, \textit{J} = 8.2 Hz, 6H,
C\textsuperscript{2',6'}H), 5.55 (s, 1H, C\textsuperscript{I}H), 2.42
(s, 9H, CH\textsubscript{3}, Ac);\textsuperscript{13}C NMR (126 MHz,
CDCl\textsubscript{3}) $\delta$ (ppm) = 193.7 (CO), 143.7
(C\textsuperscript{1'}), 137.5 (C\textsuperscript{2}H), 134.4
(C\textsuperscript{4}H), 132.7 (C\textsuperscript{6}H), 132.0
(C\textsuperscript{3',5'}H), 129.7 (C\textsuperscript{2',6'}H), 129.3
(C\textsuperscript{5}H), 128.5 (C\textsuperscript{3}), 124.7
(C\textsuperscript{1}), 121.5 (C\textsuperscript{4'}), 90.4
($\equiv$C\textsuperscript{II}--), 88.7 (--C\textsuperscript{III}$\equiv$), 56.6
(C\textsuperscript{I}H), 30.5 (CH\textsubscript{3}, Ac); FTIR
(ATR\textbf{)}: $\tilde{\nu}$ (cm\textsuperscript{-1}) = 1505 (m), 1466 (m), 1402
(w), 1351 (m), 1108 (m), 1078 (m), 1018 (w), 947 (m), 886 (w), 858 (w),
821 (m), 787 (m), 747 (w), 684 (m), 612 (m), 588 (w), 571 (w); ESI (+)
HRMS calcd for
C\textsubscript{49}H\textsubscript{34}O\textsubscript{3}S\textsubscript{3}Na:
789.1562 {[}M + Na{]}\textsuperscript{+}, found \textit{m/z} 789.1566.

\newpage
\subsection{NMR Spectra}
\newcommand{\nmrsize}{0.75}

\begin{figure}[h]
\includegraphics[width=\nmrsize\columnwidth]{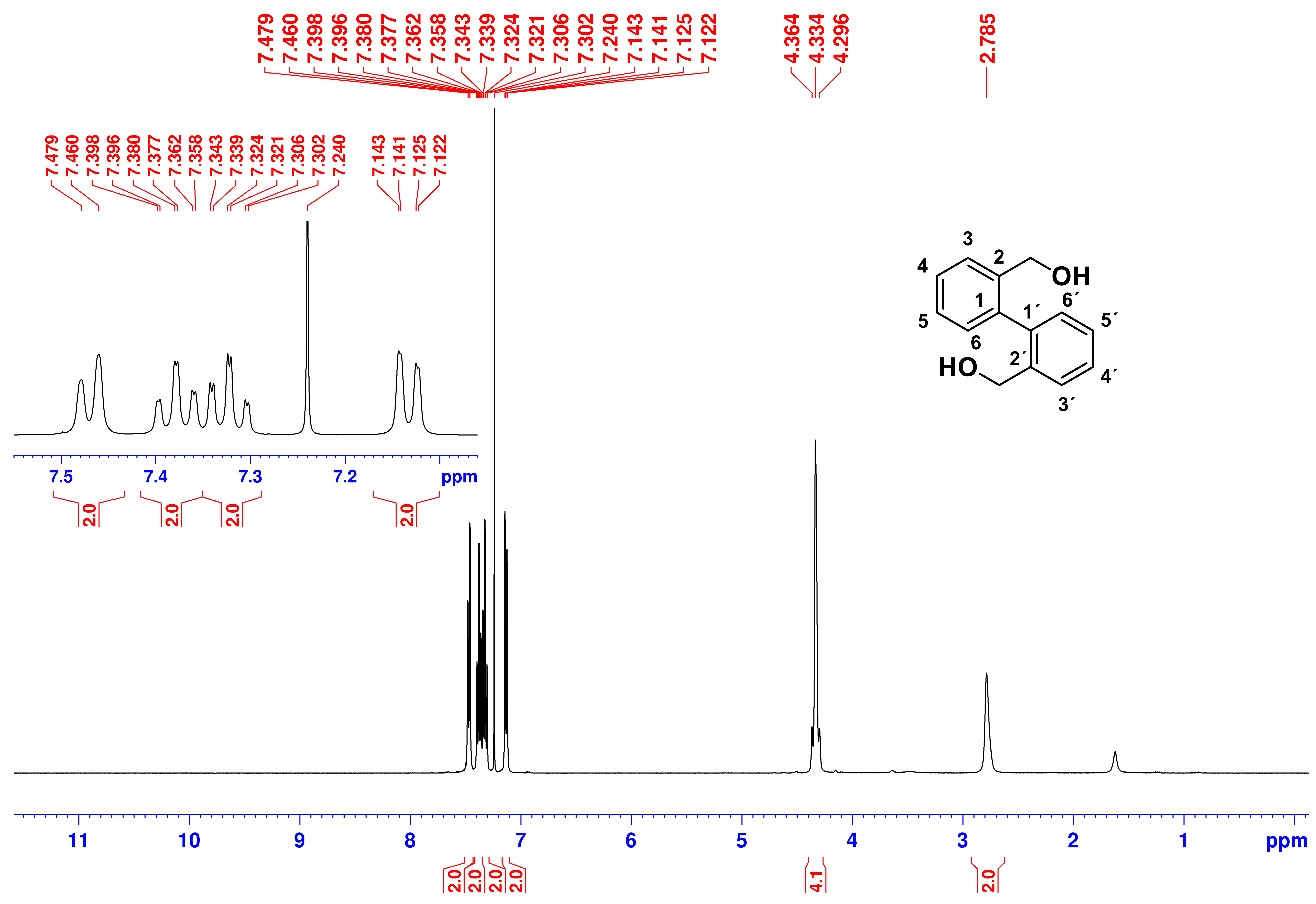}
\caption{ \textsuperscript{1}H NMR (400 MHz,
CDCl\textsubscript{3}) of compound \textbf{II}.}
\label{HNMR_II}
\end{figure}

\begin{figure}[b]
\includegraphics[width=\nmrsize\columnwidth]{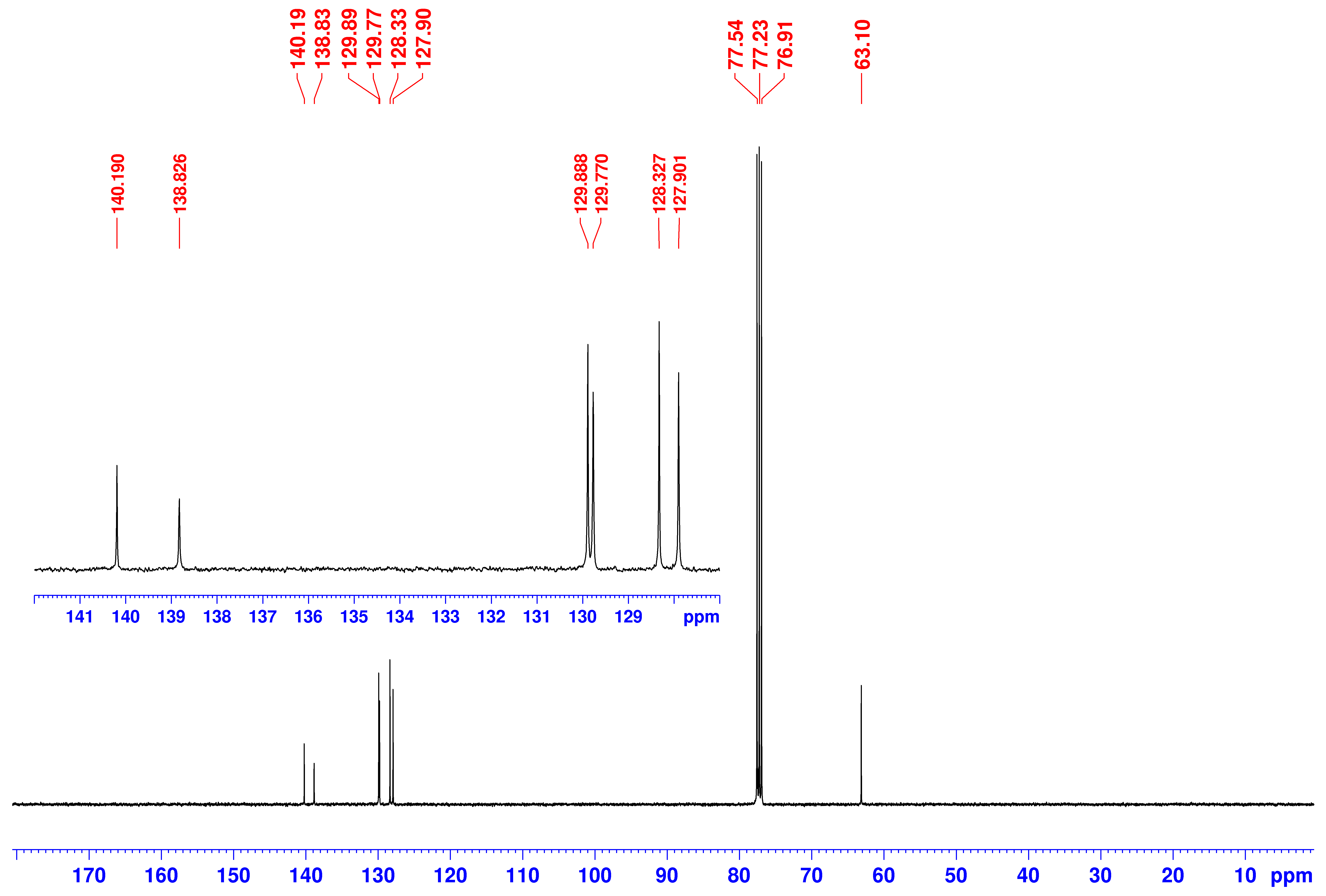}
\caption{ \textsuperscript{13}C NMR (101 MHz,
CDCl\textsubscript{3}) of compound \textbf{II}.}
\label{CNMR_II}
\end{figure}

\begin{figure}[t]
\includegraphics[width=\nmrsize\columnwidth]{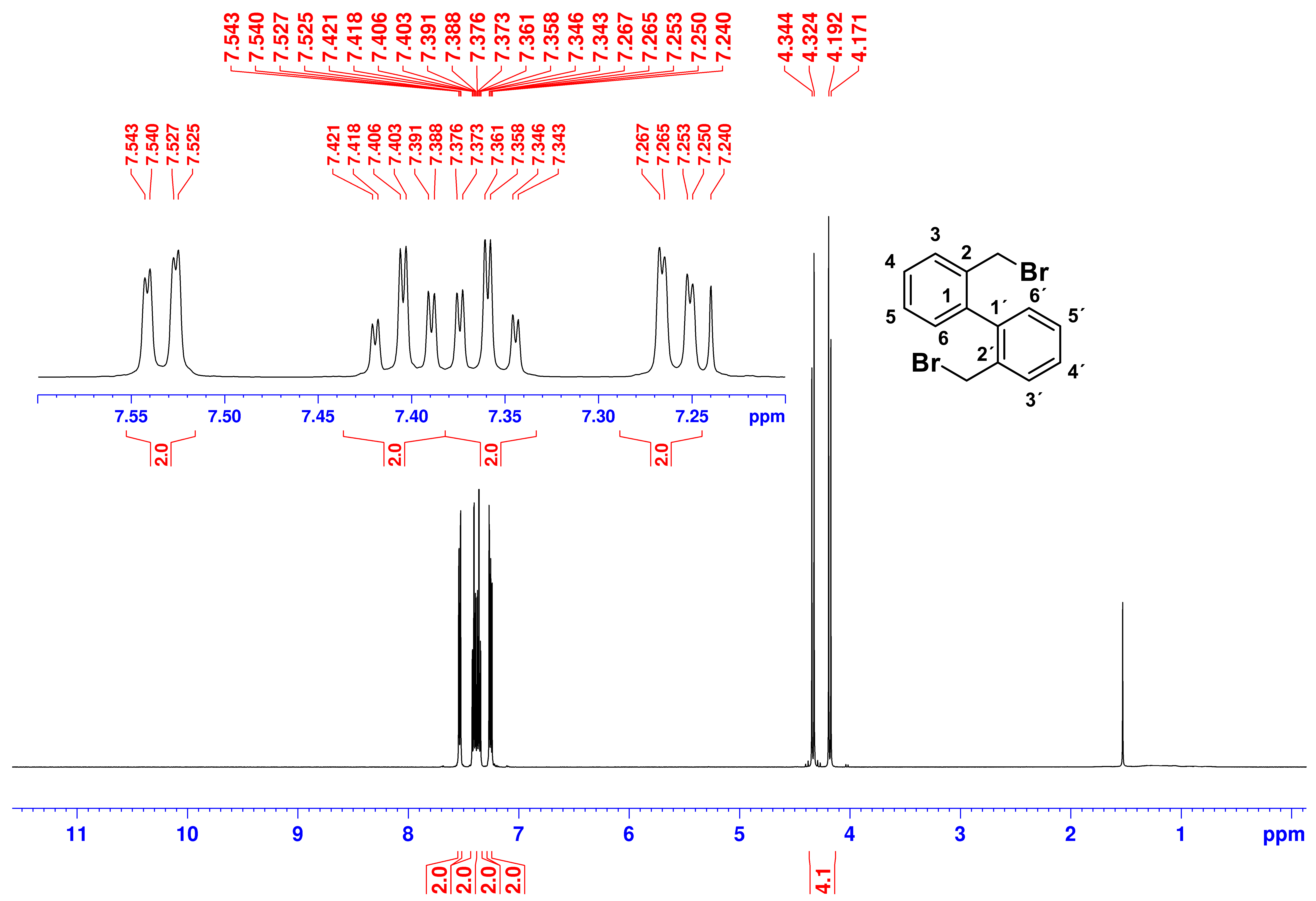}
\caption{ \textsuperscript{1}H NMR (500 MHz,
CDCl\textsubscript{3}) of compound \textbf{III}.}
\label{HNMR_III}
\end{figure}

\begin{figure}[b]
\includegraphics[width=\nmrsize\columnwidth]{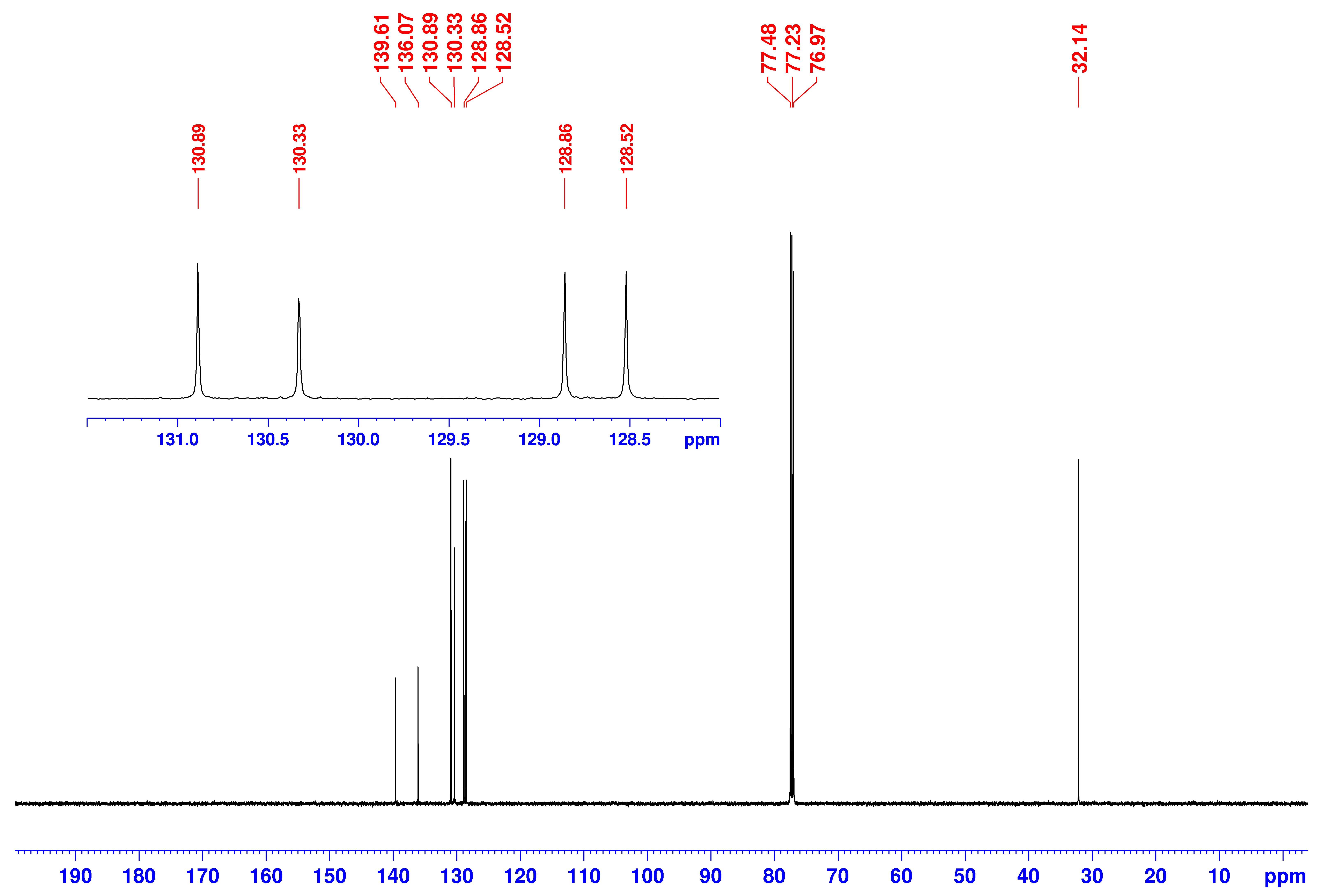}
\caption{ \textsuperscript{13}C NMR (126 MHz,
CDCl\textsubscript{3}) of compound \textbf{III}.}
\label{CNMR_III}
\end{figure}

\begin{figure}[t]
\includegraphics[width=\nmrsize\columnwidth]{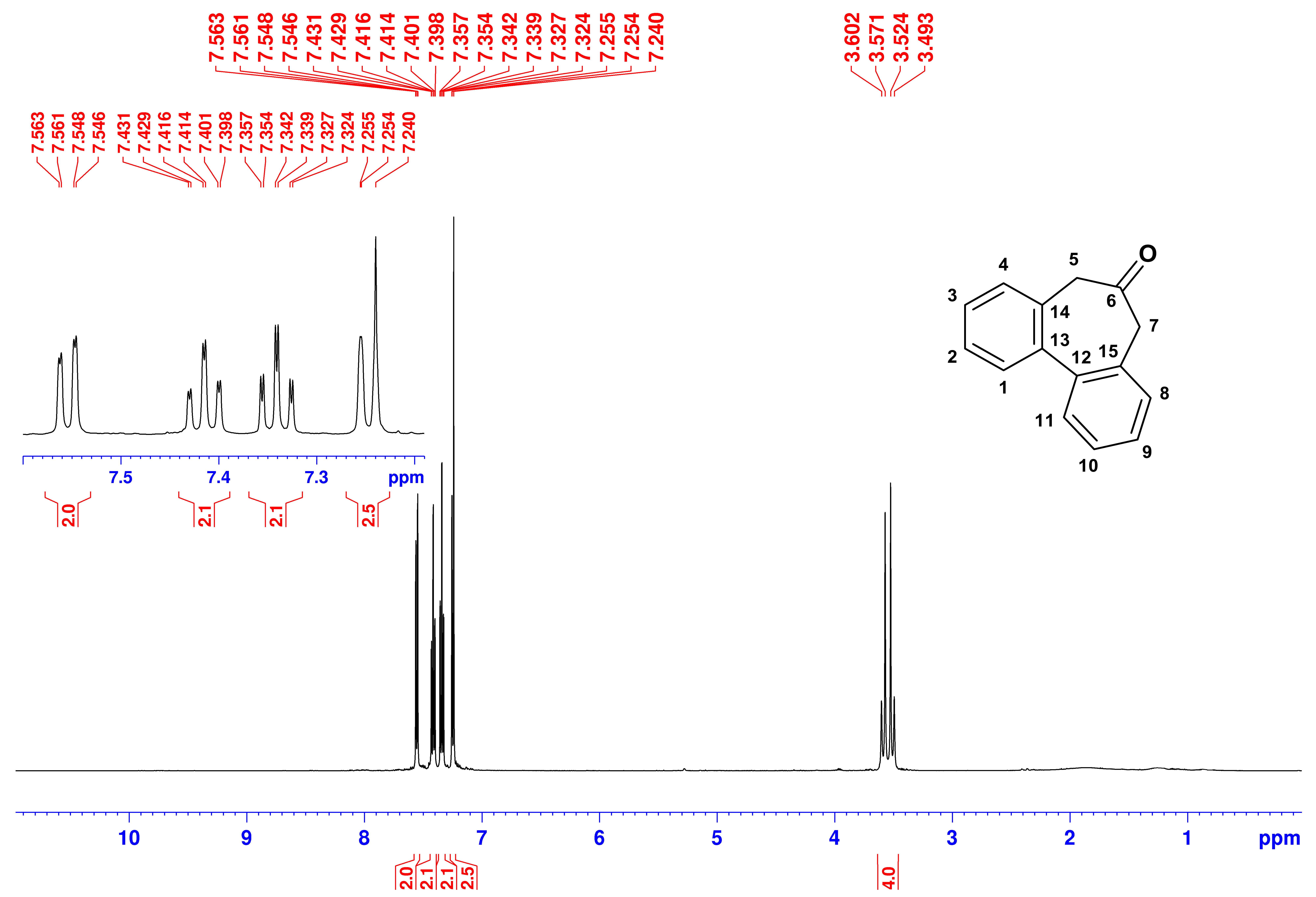}
\caption{ \textsuperscript{1}H NMR (500 MHz,
CDCl\textsubscript{3}) of compound \textbf{IV}.}
\label{HNMR_IV}
\end{figure}

\begin{figure}[b]
\includegraphics[width=\nmrsize\columnwidth]{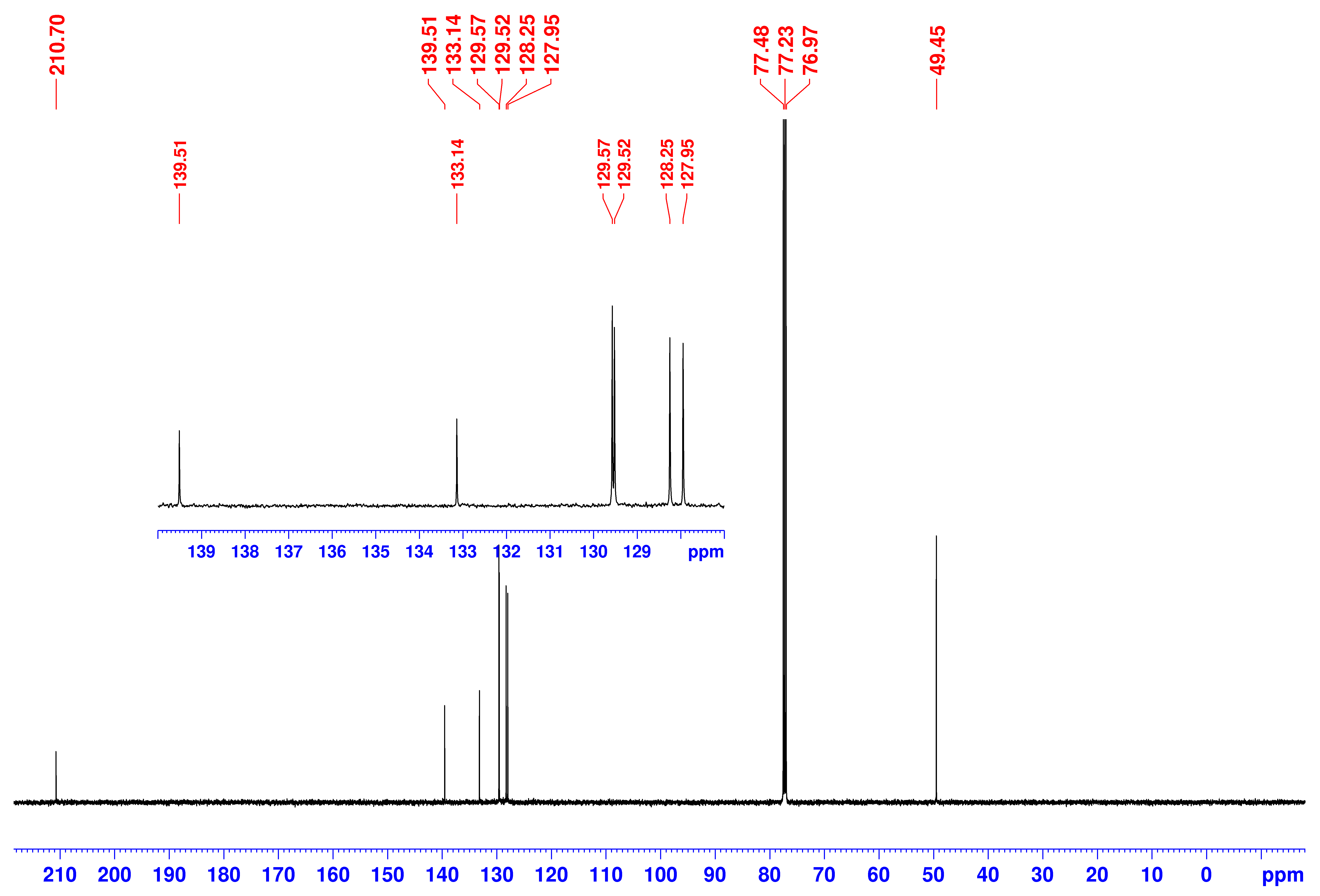}
\caption{ \textsuperscript{13}C NMR (126 MHz,
CDCl\textsubscript{3}) of compound \textbf{IV}.}
\label{CNMR_IV}
\end{figure}

\begin{figure}[t]
\includegraphics[width=\nmrsize\columnwidth]{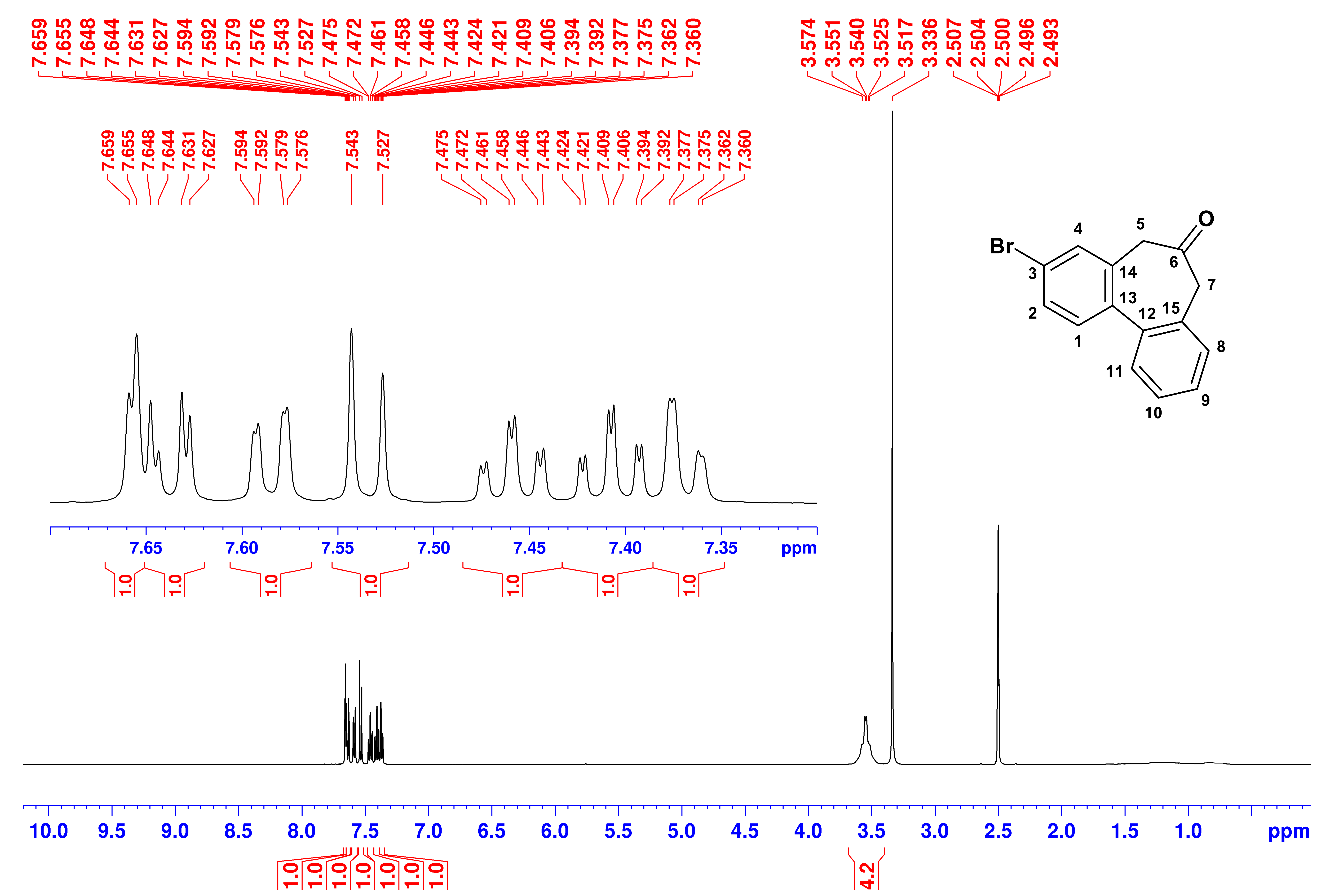}
\caption{ \textsuperscript{1}H NMR (500 MHz,
DMSO-\textit{d\textsubscript{6}}) of compound \textbf{V}.}
\label{HNMR_V}
\end{figure}

\begin{figure}[b]
\includegraphics[width=\nmrsize\columnwidth]{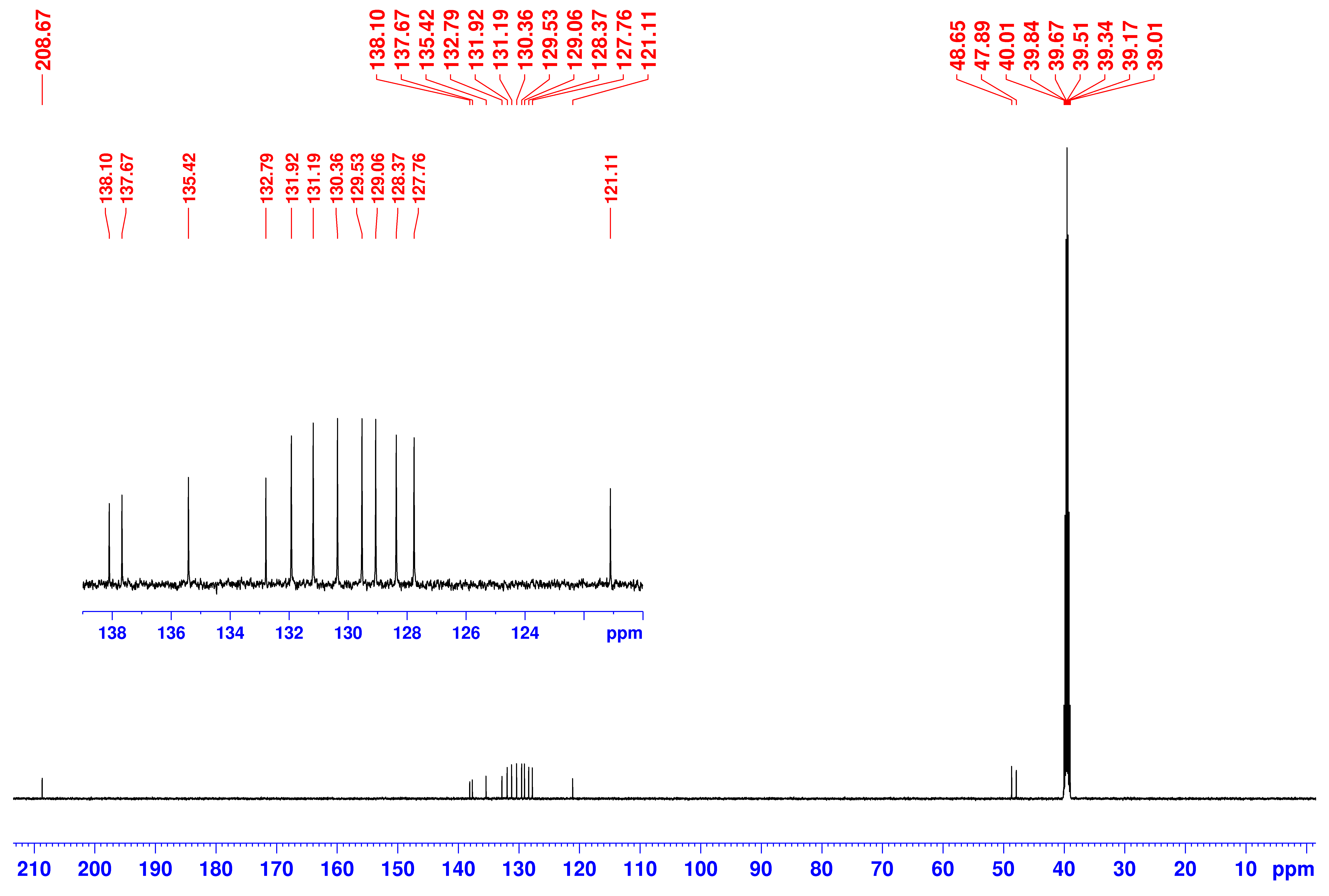}
\caption{ \textsuperscript{13}C NMR (126 MHz,
DMSO-\textit{d\textsubscript{6}}) of compound \textbf{V}.}
\label{CNMR_V}
\end{figure}

\begin{figure}[t]
\includegraphics[width=\nmrsize\columnwidth]{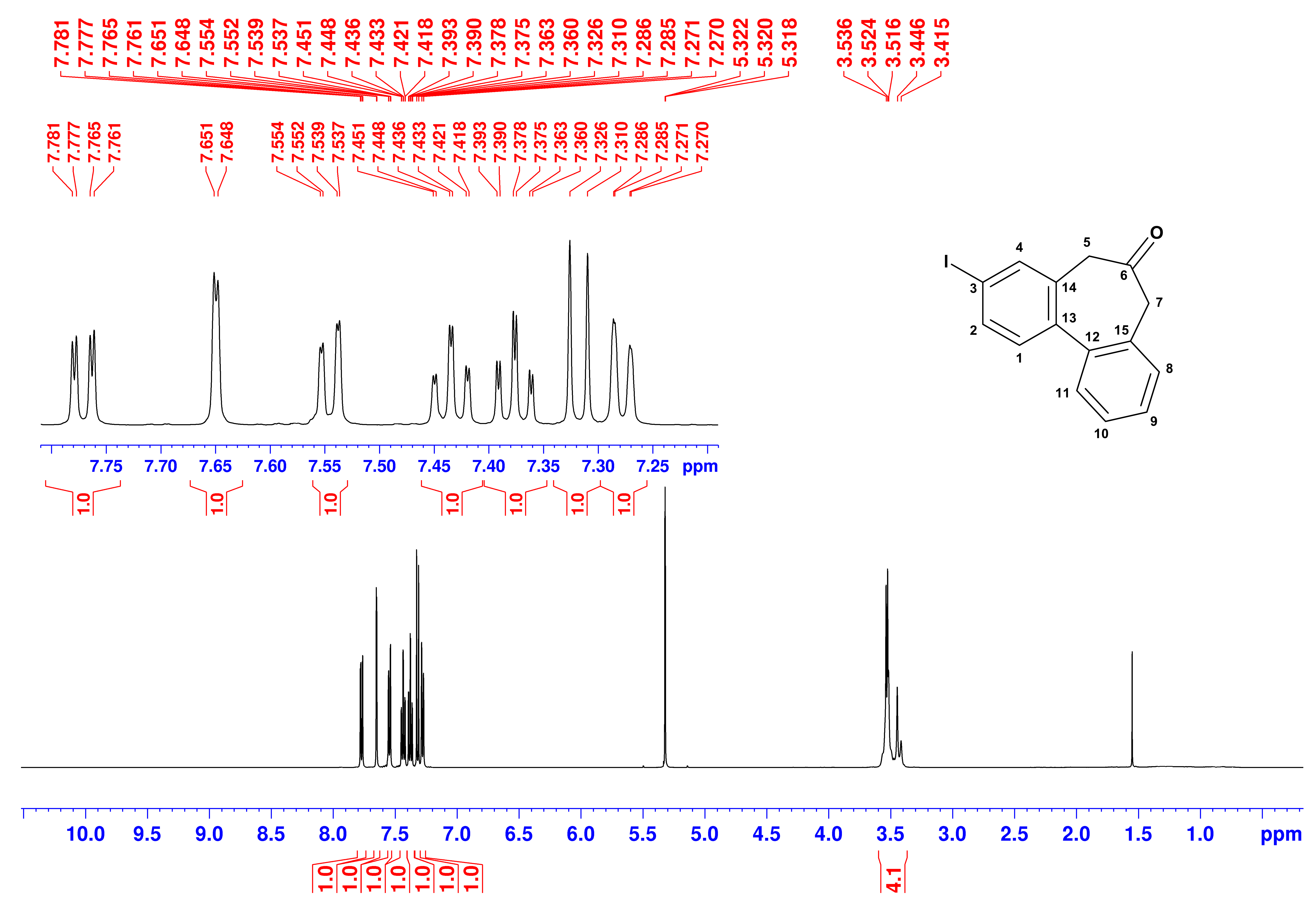}
\caption{ \textsuperscript{1}H NMR (500 MHz,
CD\textsubscript{2}Cl\textsubscript{2}) of compound \textbf{VI}.}
\label{HNMR_VI}
\end{figure}

\begin{figure}[b]
\includegraphics[width=\nmrsize\columnwidth]{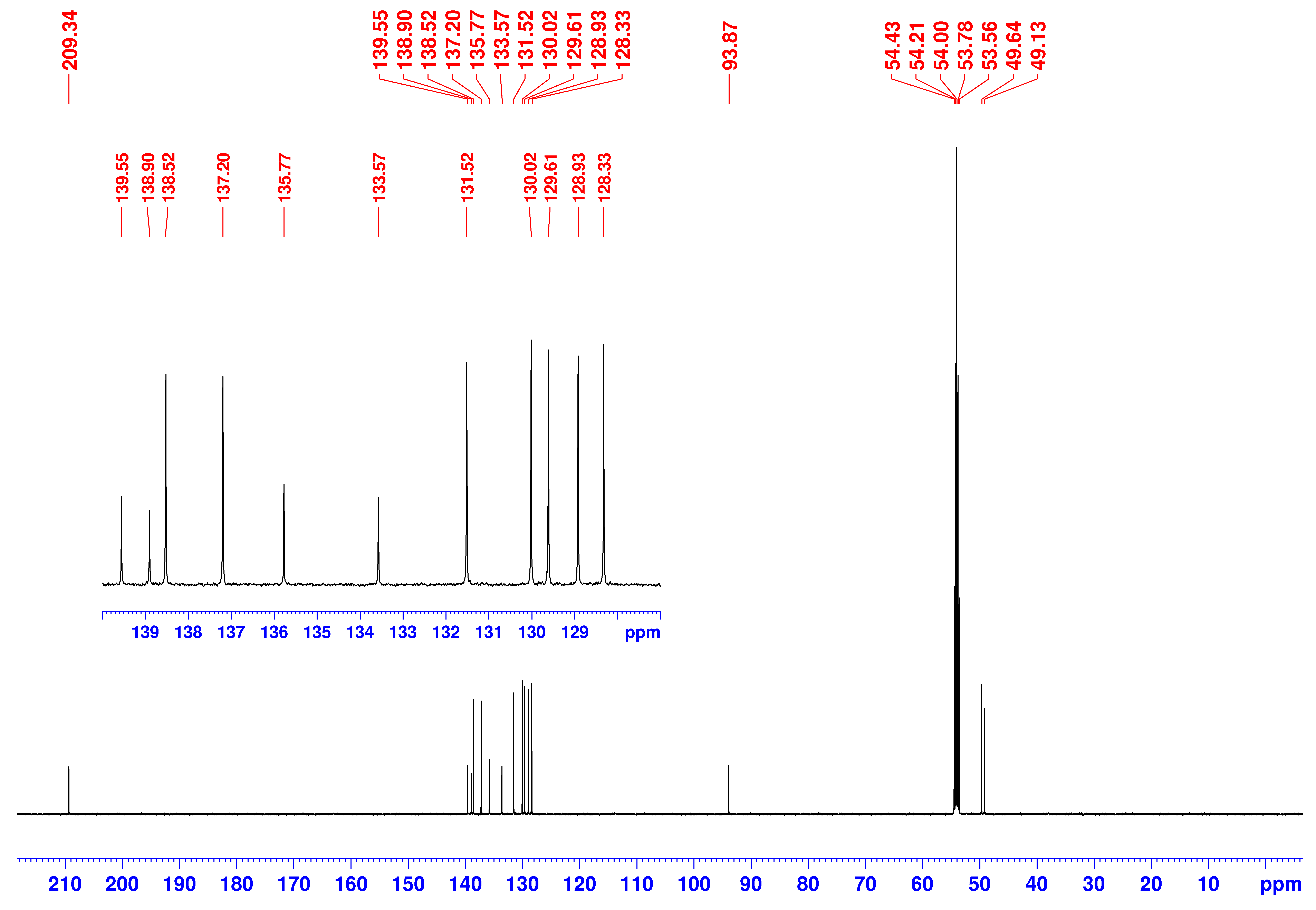}
\caption{ \textsuperscript{13}C NMR (126 MHz,
CD\textsubscript{2}Cl\textsubscript{2}) of compound \textbf{VI}.}
\label{CNMR_VI}
\end{figure}

\begin{figure}[t]
\includegraphics[width=\nmrsize\columnwidth]{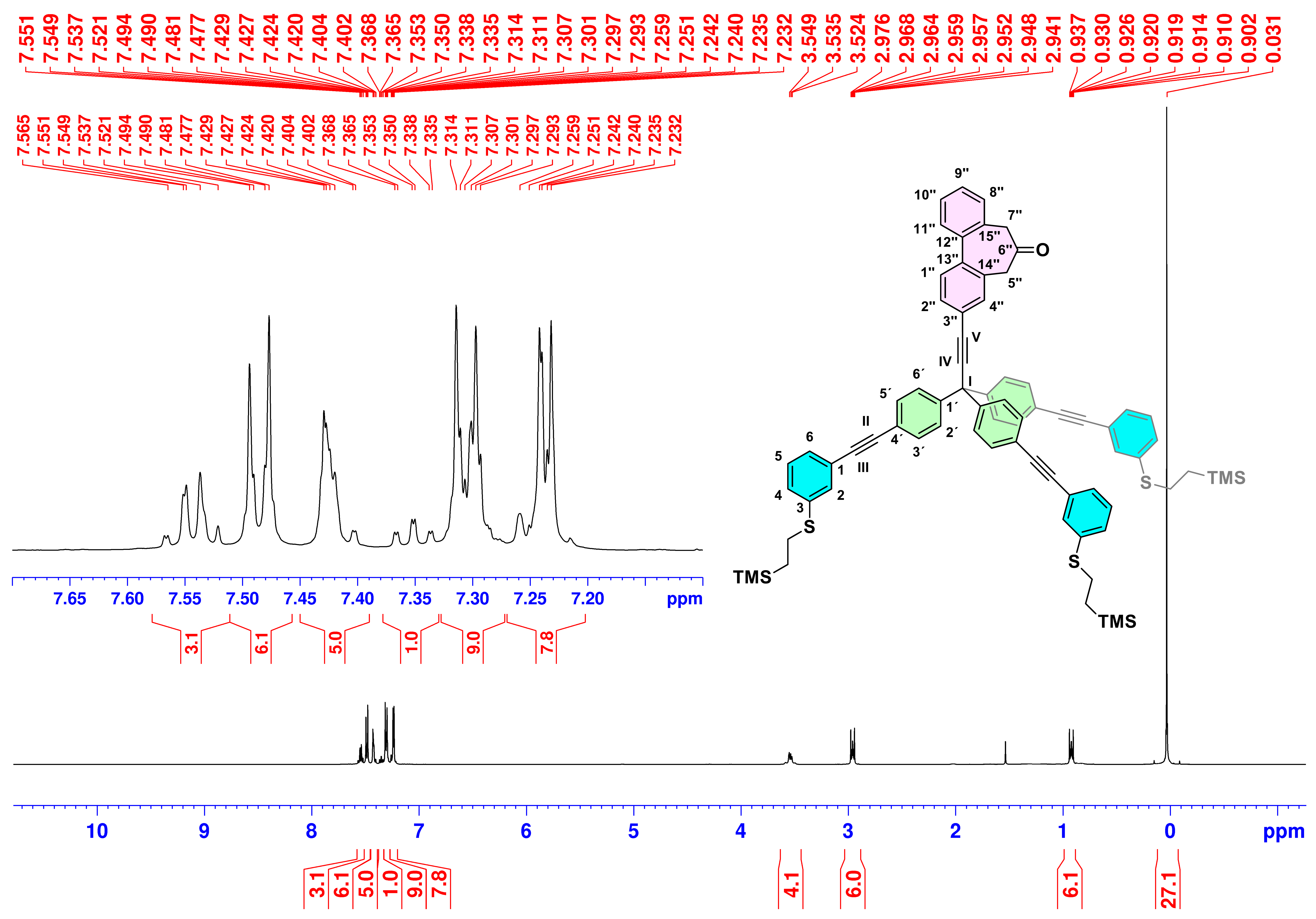}
\caption{ \textsuperscript{1}H NMR (500 MHz,
CDCl\textsubscript{3}) of compound \textbf{VIII}.}
\label{HNMR_VIII}
\end{figure}

\begin{figure}[b]
\includegraphics[width=\nmrsize\columnwidth]{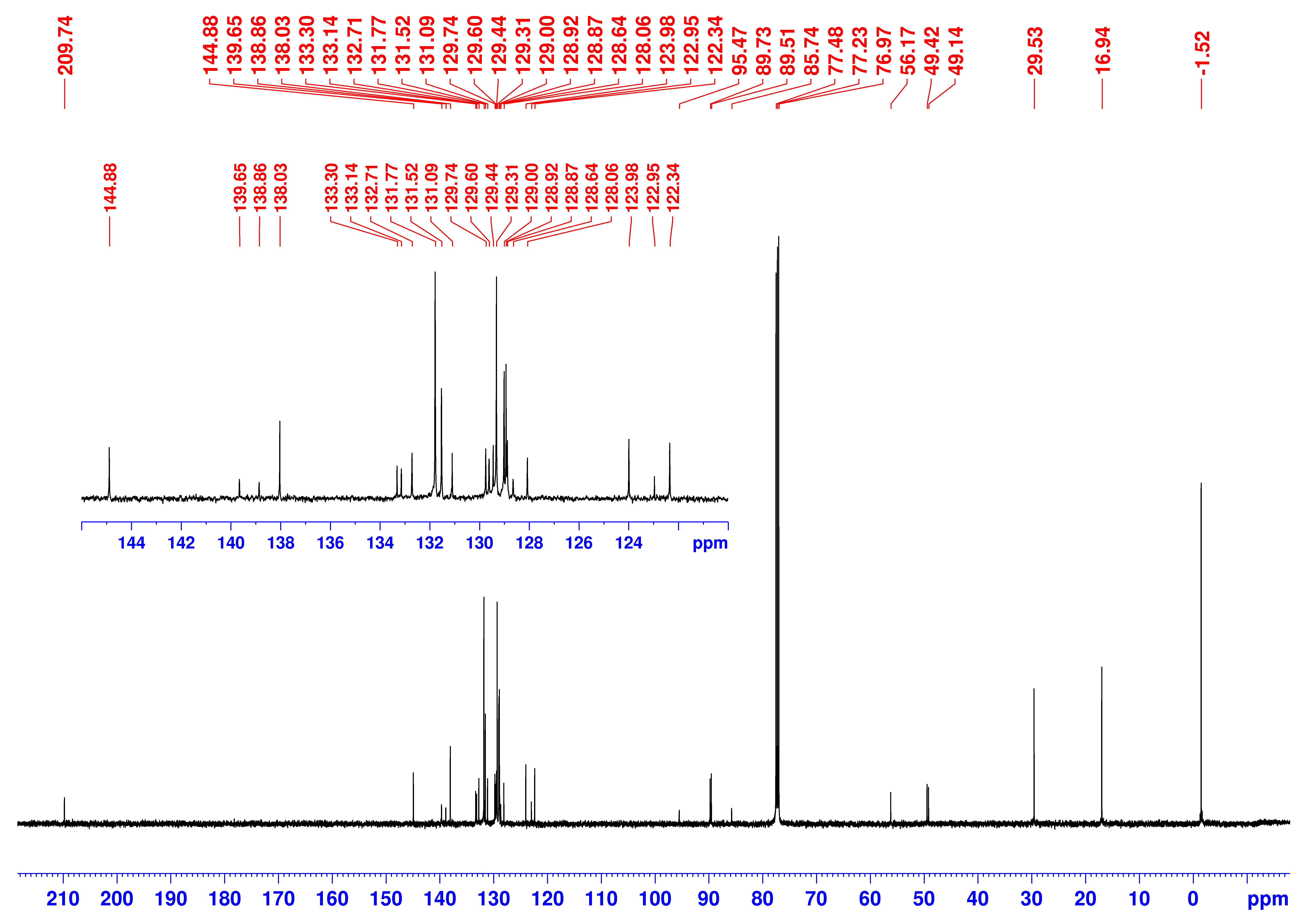}
\caption{ \textsuperscript{13}C NMR (126 MHz,
CDCl\textsubscript{3}) of compound \textbf{VIII}.}
\label{CNMR_VIII}
\end{figure}

\begin{figure}[t]
\includegraphics[width=\nmrsize\columnwidth]{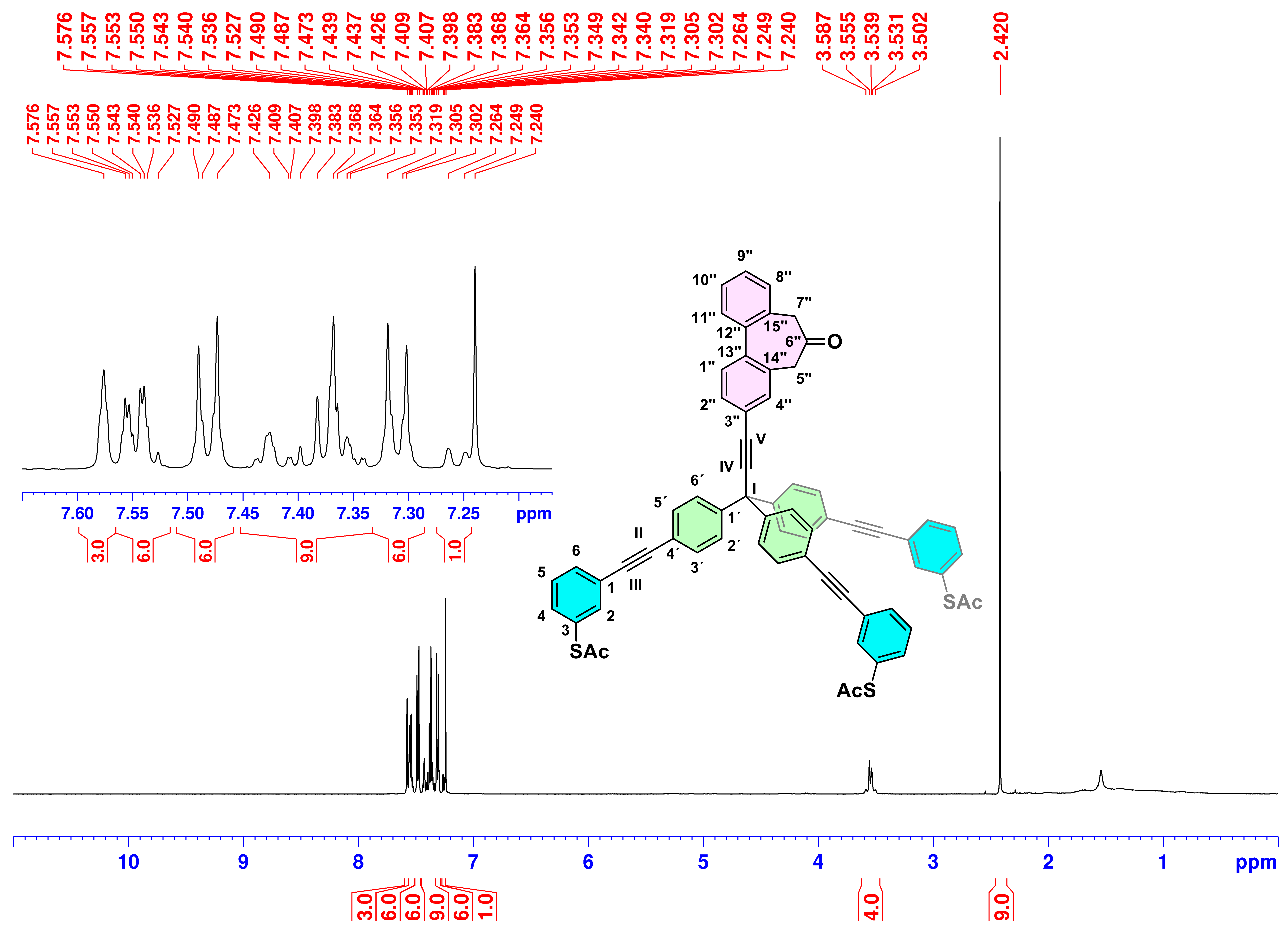}
\caption{ \textsuperscript{1}H NMR (500 MHz,
CDCl\textsubscript{3}) of compound \textbf{1}.}
\label{HNMR_1}
\end{figure}

\begin{figure}[b]
\includegraphics[width=\nmrsize\columnwidth]{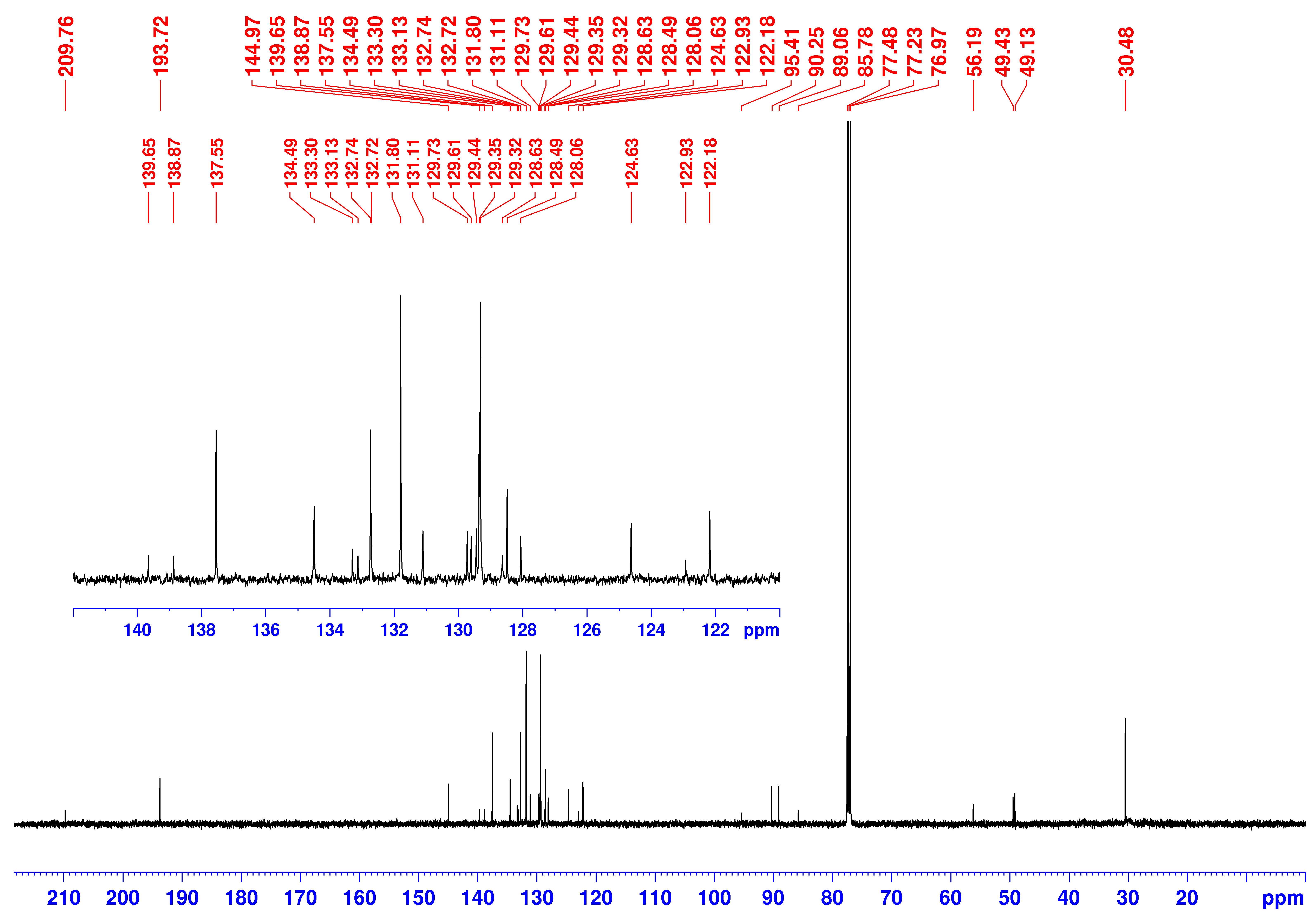}
\caption{ \textsuperscript{13}C NMR (126 MHz,
CDCl\textsubscript{3}) of compound \textbf{1}.}
\label{CNMR_1}
\end{figure}

\begin{figure}[t]
\includegraphics[width=\nmrsize\columnwidth]{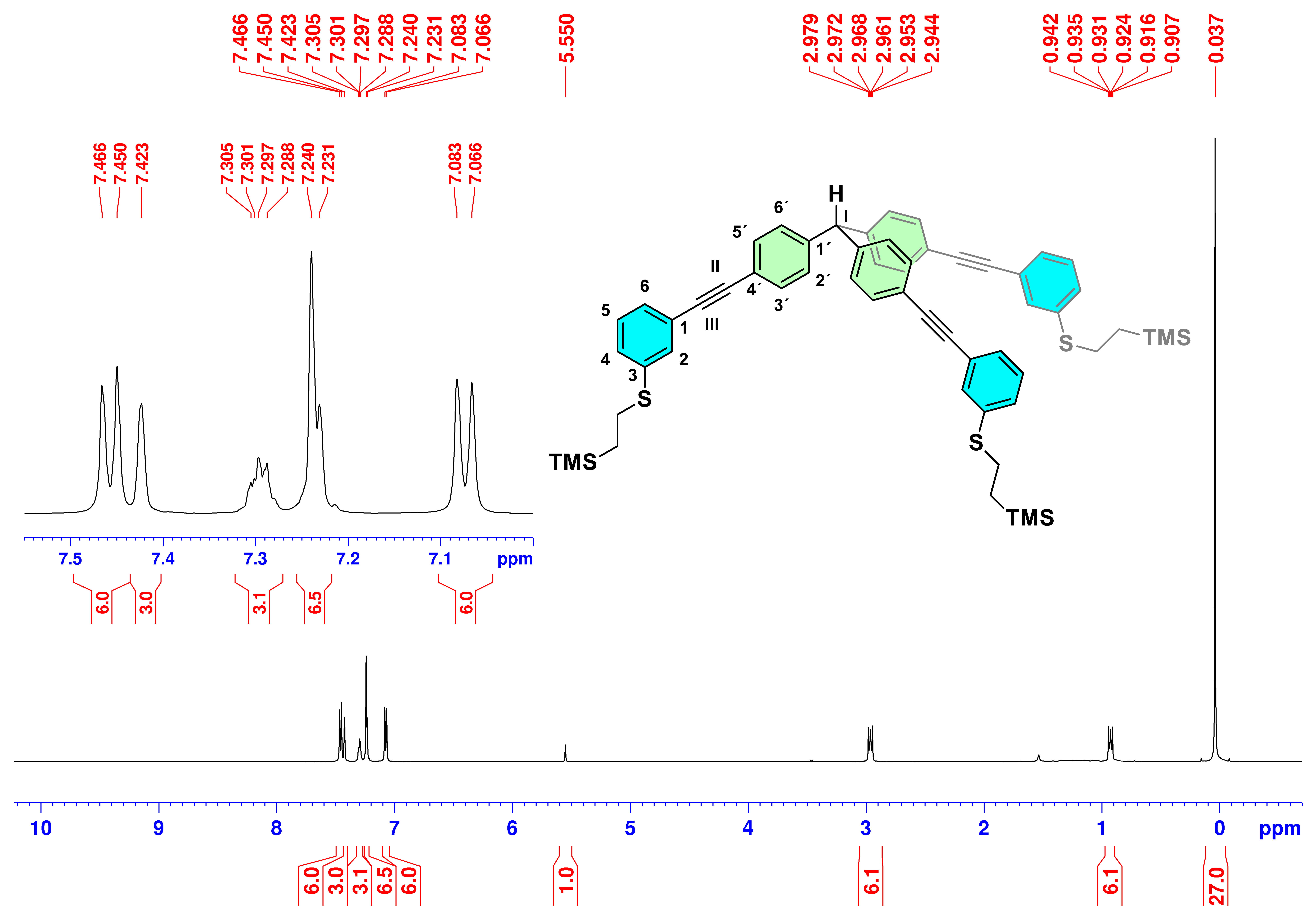}
\caption{ \textsuperscript{1}H NMR (500 MHz,
CDCl\textsubscript{3}) of compound \textbf{XI}.}
\label{HNMR_XI}
\end{figure}

\begin{figure}[b]
\includegraphics[width=\nmrsize\columnwidth]{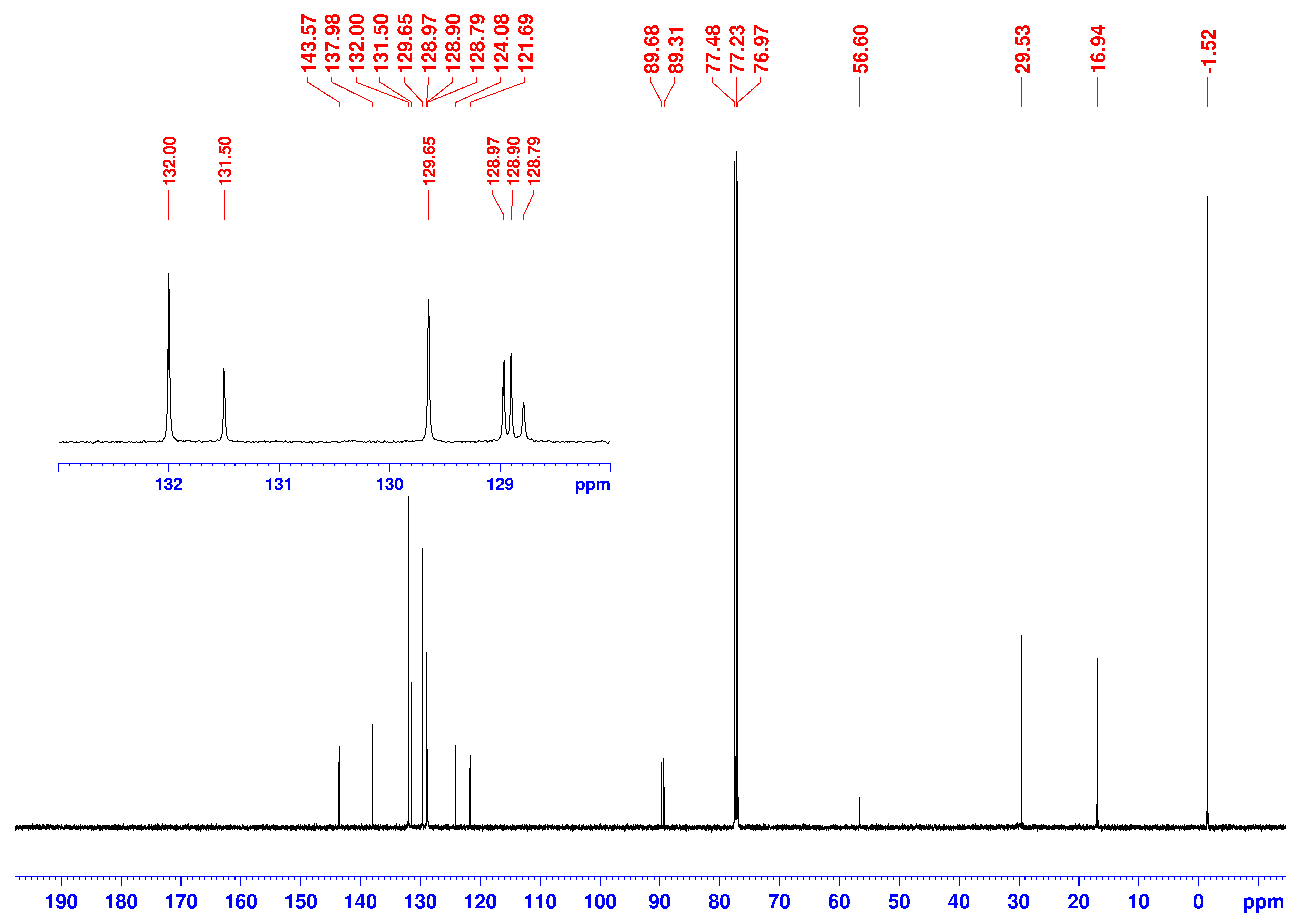}
\caption{ \textsuperscript{13}C NMR (126 MHz,
CDCl\textsubscript{3}) of compound \textbf{XI}.}
\label{CNMR_XI}
\end{figure}

\begin{figure}[t]
\includegraphics[width=\nmrsize\columnwidth]{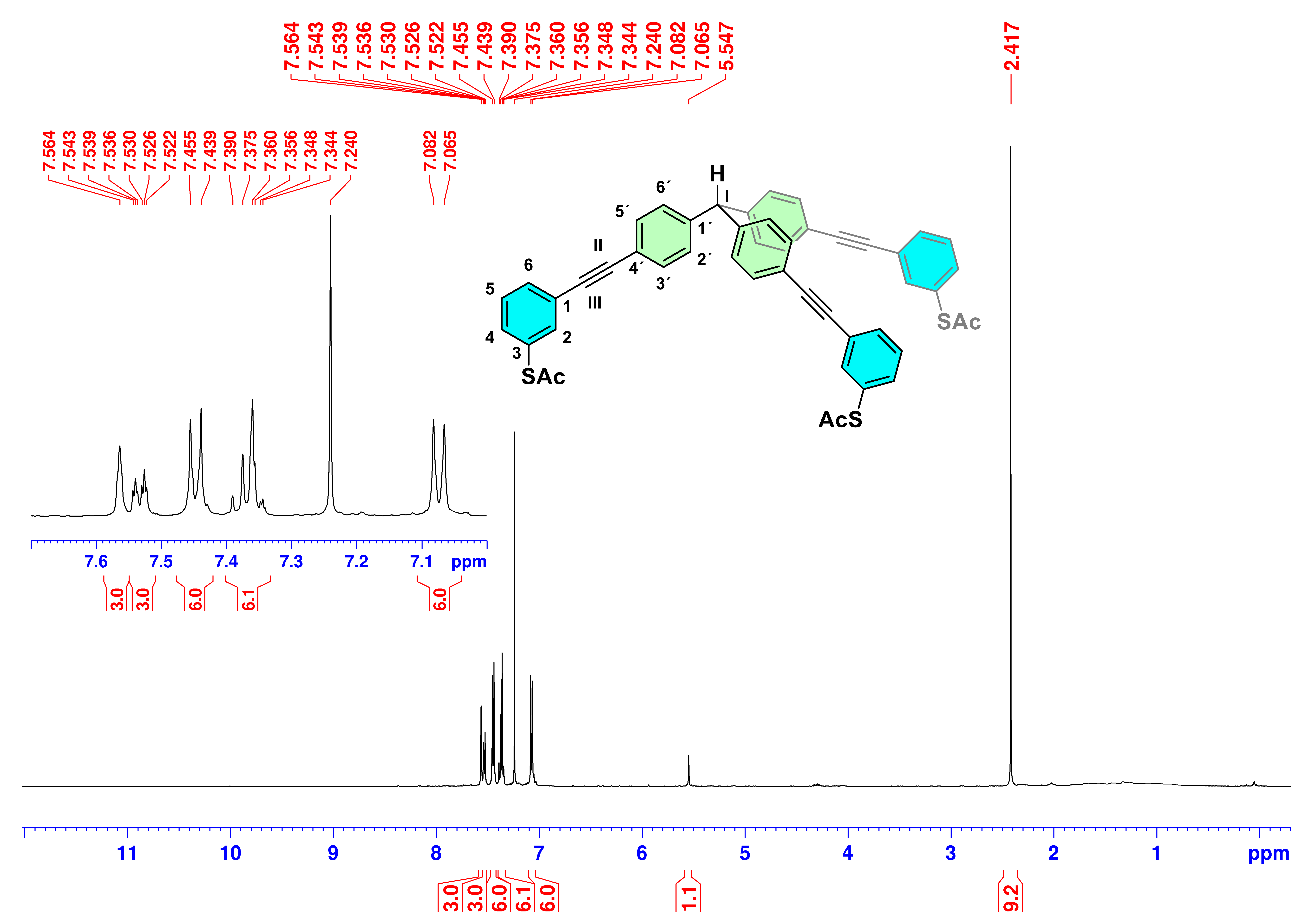}
\caption{ \textsuperscript{1}H NMR (500 MHz,
CDCl\textsubscript{3}) of compound \textbf{2}.}
\label{HNMR_2}
\end{figure}

\begin{figure}[b]
\includegraphics[width=\nmrsize\columnwidth]{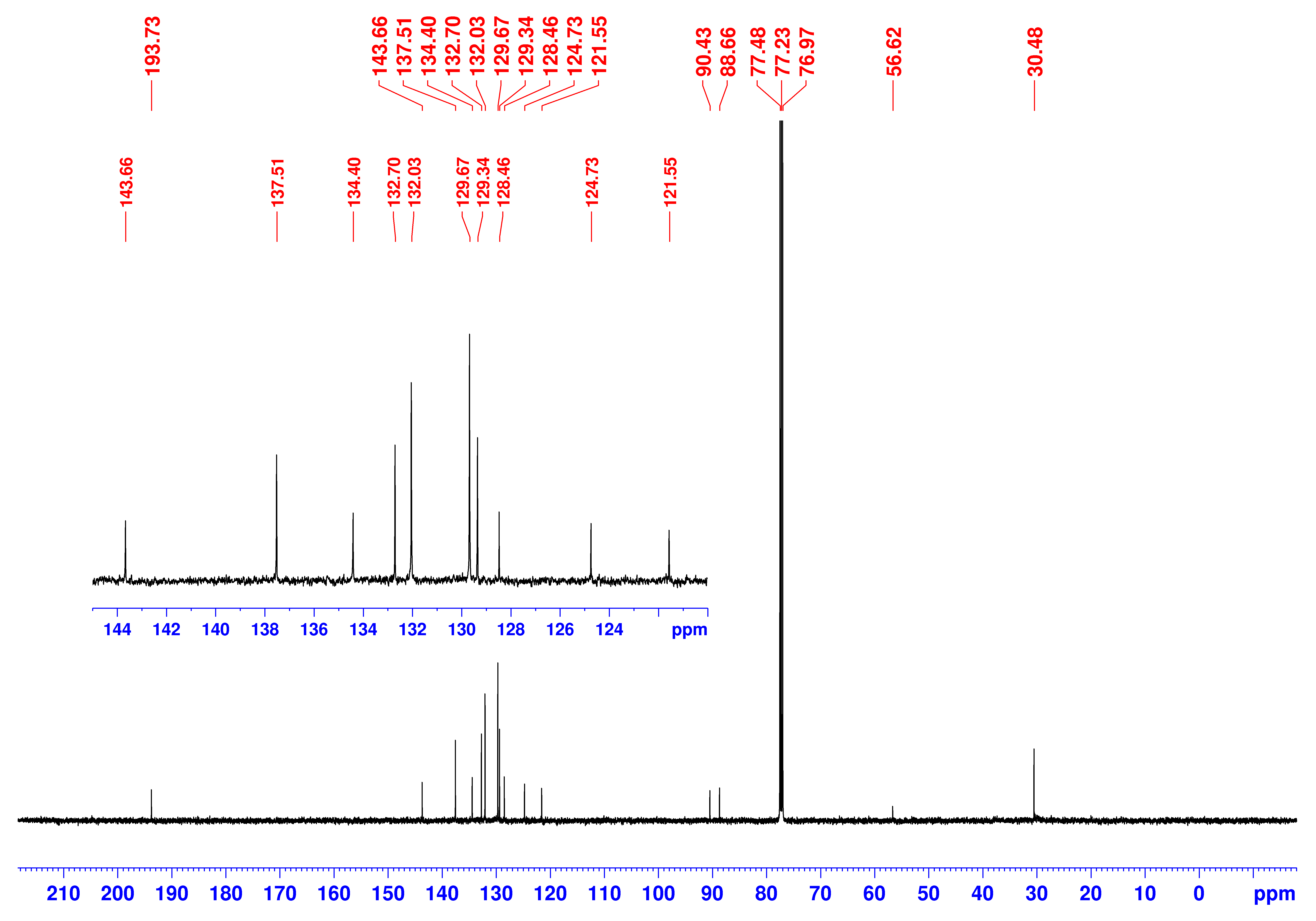}
\caption{ \textsuperscript{13}C NMR (126 MHz,
CDCl\textsubscript{3}) of compound \textbf{2}.}
\label{CNMR_2}
\end{figure}

\clearpage

\section{Additional STM experiments}

\begin{figure}[h]
	\centering
	\includegraphics[width=11.5cm]{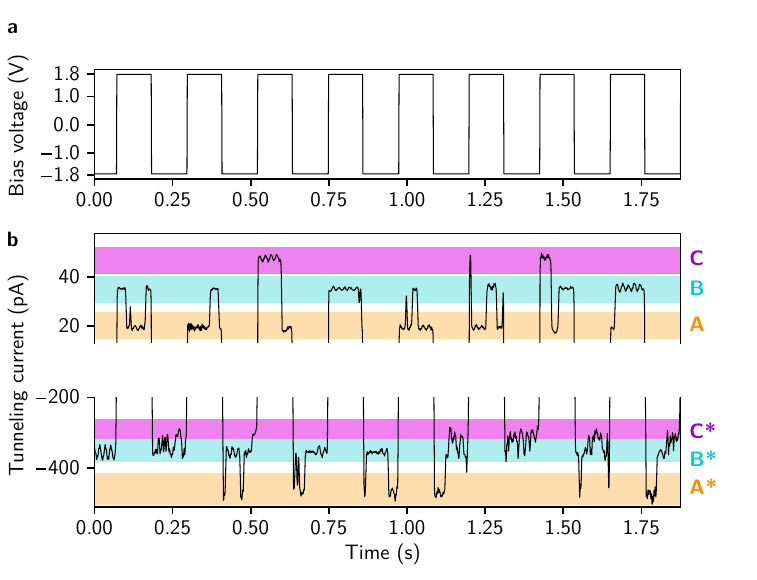}
	\caption{\textbf{Correlation of states observed at different bias polarity. a} Tunneling voltage applied to the sample. \textbf{b} Tunneling current. The full time trace comprises 99 inversions of the polarity of the voltage. For 87 inversions, the current levels switch according to the correspondence A*/B*/C* $\equiv$ A/B/C as indicated by color.}
	\label{SI_STM_corr}
\end{figure}

\begin{figure}[h]
	\centering
	\includegraphics[width=0.7\columnwidth]{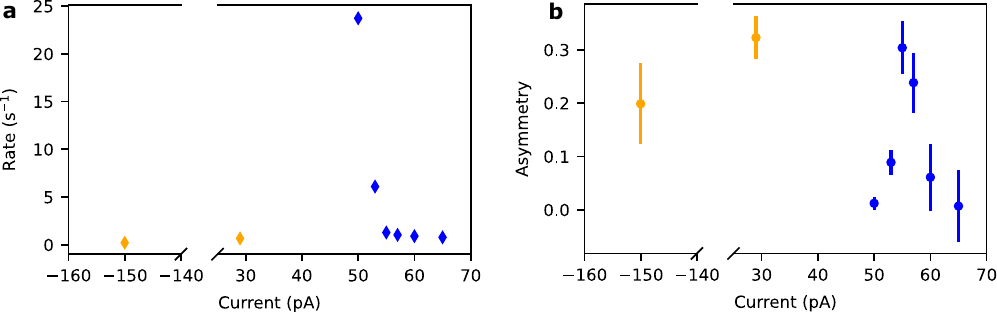}
	\caption{\textbf{Current dependency of rotational switching.} \textbf{a} Number of transitions per second and \textbf{b} asymmetry between the two switching directions as a function of the average tunneling current (set by the \textit{z} position of the STM tip) recorded at a bias voltage of 1.83 V (blue dots) and 1.85 V and -1.85 V (orange dots). Blue and orange markers each correspond to one fixed position of the STM tip above a molecule.}
	\label{SI_STM_deps}
\end{figure}

\clearpage
\bibliography{short_refs.bib}